\documentclass[Journal,comsoc]{IEEEtran}
 \def\doublecolumn{1}

\usepackage[T1]{fontenc} 
\usepackage{amsmath}
\usepackage[cmintegrals]{newtxmath}
\usepackage{bm} 

\usepackage{cite}  
\usepackage{tikz}
\usetikzlibrary{graphs} 
\usepackage{multirow} 
\usepackage{colortbl} 
\usepackage{booktabs} 

\newcommand\VRule[1][\arrayrulewidth]{\vrule width #1}

\usepackage{etoolbox}
\makeatletter
\patchcmd{\@begintheorem}{\textit}{\textbf}{}{}
\makeatother

\newtheorem{theorem}{Theorem}
\newtheorem{remark}{Remark}
\newtheorem{lemma}{Lemma}
\newtheorem{definition}{Definition}

\newtheorem{proposition}{Proposition}

\newtheorem{observation}{Observation}

\newcommand{\mais}{\mathsf{mais}}
\newcommand{\minrank}{\mathsf{minrk}_2 }

\newcommand{\od}[2]{d^+_{#1}\!(#2)}
\newcommand{\on}[2]{N^+_{#1}\!(#2)}

\begin{document}

\title{Optimal Finite-Length and Asymptotic\\ Index Codes for Five or Fewer Receivers}
\author{Lawrence Ong 
\thanks{This research was supported under Australian Research Council's Future Fellowship (project number FT140100219) and Discovery Project (project number DP150100903)  funding schemes.

  This paper was presented in part at the 2014 International Symposium on Network Coding and at the 2014 IEEE International Symposium on Information Theory.



}
}
\maketitle

\begin{abstract}
Index coding models broadcast networks in which a sender sends different messages to different receivers simultaneously, where each receiver may know some of the messages a priori. The aim is to find the minimum (normalised) index codelength that the sender sends.
This paper considers unicast index coding, where each receiver requests exactly one message, and each message is requested by exactly one receiver.  Each unicast index-coding instances can be fully described by a directed graph and vice versa, where each vertex corresponds to one receiver. For any directed graph representing a unicast index-coding instance, we show that if a maximum acyclic induced subgraph (MAIS) is obtained by removing two or fewer vertices from the graph, then the minimum index codelength equals the number of vertices in the MAIS, and linear codes are optimal for the corresponding index-coding instance. Using this result, we solved all unicast index-coding instances with up to five receivers, which correspond to all graphs with up to five vertices. For 9819 non-isomorphic graphs among all graphs up to five vertices, we obtained the minimum index codelength for all message alphabet sizes; for the remaining 28 graphs, we obtained the minimum index codelength if the message alphabet size is $k^2$ for any positive integer $k$. This work complements the result by Arbabjolfaei et al.\ (ISIT 2013), who solved all unicast index-coding instances with up to five receivers in the asymptotic regime, where the message alphabet size tends to infinity.
\end{abstract}

\begin{IEEEkeywords}
Index coding, broadcast with side information, graph theory, finite-length codes
\end{IEEEkeywords}

\section{Introduction} \label{section:introduction}

Index coding~\cite{birkkol2006,baryossefbirk11} studies noiseless one-hop broadcast networks, with one sender and multiple receivers. The sender has a set of messages, and each receiver wants a message subset, while knowing another message subset a priori. To this end, the sender encodes the messages into an index codeword and presents the codeword to all the receivers. The index codeword enables each receiver to decode its requested message subset. In majority of the work on index coding, the aim is to minimise the normalised index codelength. Index coding have been receiving much attention lately, partly due to its equivalence to network coding~\cite{rouayhebsprintsongeorghiades10,effrosrouayheblangberg15}.

To date, different index-code construction techniques have been proposed~\cite{baryossefbirk11, neelytehranizhang13, shanmugamdimakislangberg13, arbabjolfaei13, unalwagner16, thapaongjohnson17it}, but none are optimal in general. Among them, composite coding~\cite{arbabjolfaei13} have been shown to achieve the optimal (i.e., minimum) normalised codelength asymptotically (as the message size tends to infinity) for \textit{unicast} index coding---where each receiver requests only one message, and each message is requested by only one receiver---if there are five or fewer receivers. In a more general setting (not necessarily unicast), Unal and Wagner~\cite{unalwagner16} solved all index-coding instances with three receivers in the asymptotic regime.

In this paper, we consider unicast index coding where the message alphabet size is finite, and derive the optimal index codelength for all instances in this class with five or fewer receivers. Our result uses combinatorics and is derived based on our graph-theoretic result that shows for any directed graph in which no two cycles are disjoint, if the a maximum acyclic induced subgraph (MAIS) is obtained by removing two or fewer vertices from the graph, then there must exist a subgraph of a certain form (see Figure~\ref{fig:three-cycles}). We incidentally showed that linear index codes are optimal for all unicast index-coding instances with up to and including five receivers. 

The rest of the paper is organised as follows: We formally define unicast index coding in Section~\ref{section:index-coding}. We survey existing results and summarise our contributions in this paper in Section~\ref{sec:main}. We present our results in two parts: Section~\ref{section:result-1} for graphs with specific MAIS values, and Section~\ref{sec:result-2} for graphs with five or fewer vertices. 

\section{Index Coding: Definition and Notation} \label{section:index-coding}


\subsection{Unicast index coding and information-flow graph}
A unicast index-coding instance consists of a single sender and multiple receivers $[n] \triangleq \{1,2,\dotsc,n\}$. The sender has $n$~messages, denoted by $\bm{X} = [X_1\, X_2 \,\dotsm X_n]$, where $X_i$ for each $i \in [n]$ is independent and uniformly distributed over a finite alphabet $\mathcal{X}$. For a subset of integers $I = \{i_1, i_2, \dotsc, i_{|I|}\}$ where $i_1 < i_2 <  \dotsm < i_{|I|}$, let $\bm{X}_I \triangleq [X_{i_1} X_{i_2} \dotsm X_{i_{|I|}}]$.  Each receiver~$i\in [n]$ has a priori knowledge of $\bm{X}_{K_i}$ for some $K_i \subseteq [n] \setminus \{i\}$, and needs to decode $X_i$. The sender is to encode $\bm{X}$ and present the codeword to all receivers, such that
each receiver~$i \in [n]$ uses the codeword and the messages $\bm{X}_{K_i}$  it already knows to decode  $\bm{X}_i$.
The aim is for the sender to minimise its transmitted information through the channel so that each receiver can recover its requested message.
Each unicast index-coding instance is completely defined by $\{K_i\}_{i=1}^n$ and $|\mathcal{X}|$.

A unicast index-coding instance can be represented by a directed graph $G$ with a set of vertices, $V(G)=[n]$, and a set of arcs, $A(G)$. An arc from vertex~$i$ to vertex~$j$, denoted by $(i \rightarrow j) \in A(G)$, exists if and only if receiver~$i$ knows $x_j$ a priori. This means the side information of receiver~$i$ is $K_i = \on{G}{i}$, where $\on{G}{i}$ is the out-neighbourhood of vertex~$i$ in $G$. By definition, there is no self loop or parallel arcs. 
This representation is known as the \emph{side-information graph}~\cite{baryossefbirk11}. 

\subsection{Index codes}
Let $\od{G}{i}$ denote the out-degree of the vertex~$i$ in graph $G$. 
\begin{definition}
An index code $(\phi, \{\psi_i\})$ for an index-coding instance $G$ with message alphabet $\mathcal{X}$ consists of the following:
\begin{enumerate}
\item A sender encoding function $\phi : \mathcal{X}^n \mapsto \mathcal{Y}^p$, for some finite alphabet $\mathcal{Y}$ and a positive integer~$p \in \mathbb{Z}^+$; and
\item A receiver decoding function $\psi_i: \mathcal{Y}^p \times \mathcal{X}^{\od{G}{i}} \mapsto \mathcal{X}$, each for a receiver~$i \in [n]$, 
\end{enumerate}
such that $X_i = \psi_i (\phi(\bm{X}),\bm{X}_{\on{G}{i}})$.
\end{definition}


\subsection{Asymptotic vs finite-length index codes}

Fix $G$ and $|\mathcal{X}| = m^t$, for some integers $m \geq 2$ and $t \geq 1$. The index codelength, in bits, for an index code $(\phi,\{\psi_i\})$ is $p \log_2|\mathcal{Y}|$ bits (need not be an integer); the \textit{normalised} codelength, or commonly referred to as the broadcast rate, is denoted as
\begin{equation}
\ell_{m^t}(G)  \triangleq \frac{p \log_2 |\mathcal{Y}|}{\log_2|\mathcal{X}|} = p \log_{|\mathcal{X}|} |\mathcal{Y}|
\end{equation}
transmitted bits per message bit per receiver. A broadcast rate is said to be \textit{achievable} if there exists an index code of that rate.
For the rest of this paper, unless otherwise stated, we refer to the normalised codelength simply as codelength.

So, for a given message alphabet size $|\mathcal{X}| = m^t$, the minimum\footnote{The minimum exists because $1 \leq \ell_{m^t}(G) \leq n$, where the lower bound follows from each receiver having to decode one message $X_i \in \mathcal{X}$ (which is independent of all its side information) from the codeword $\phi(\bm{X}) \in \mathcal{Y}^p$; the upper bound is obtained by sending all messages uncoded $\phi(\bm{X}) = \bm{X}$. So, $r_{m^t}(G)$ is obtained by minimising $\ell_{m^t}(G)$ with a finite combinations of $|\mathcal{Y}| \in \{2, \dotsc, |\mathcal{X}|^n\}$ and $p \in [\lceil n \log_2|\mathcal{X}|\rceil]$.} codelength, over all possible index codes, is given by
\begin{equation}
r_{m^t}(G) \triangleq \min_{\phi,\{\psi_i\}} \ell_{m^t}(G). \label{eq:broadcast-rate}
\end{equation}

Furthermore, we define the \textit{optimal} index codelength (or the optimal broadcast rate) for an index-coding instance $G$, over all message alphabet sizes and all index codes, as
\begin{equation}
r(G) \triangleq \inf_{m,t} \min_{\phi,\{\psi_i\}} \ell_{m^t}(G) =\inf_{m,t} r_{m^t}(G). \label{eq:optimal-broadcast-rate}
\end{equation}
The optimal index codelength is also known as the \textit{beta capacity} $\beta(G)$.

We now show that the optimal index codelength can be obtained by taking the limit of $r_{m^t}(G)$ as $t \rightarrow \infty$ for any $m$, stated in the following proposition:
\begin{proposition} \label{prop:equivalent-beta}
For any $m$,
\begin{equation}
\lim_{t \rightarrow \infty} r_{m^t}(G) = \inf_t r_{m^t}(G)= r(G). \label{eq:optimal-broadcast-rate-2}
\end{equation}
\end{proposition}

\begin{IEEEproof}
We fix $m$ and vary $t$. Denote the absolute (not normalised) minimum codelength by $r^\text{b}_{m^t} \triangleq \min_{\phi,\{\psi_i\}} p \log_2|\mathcal{Y}|$ bits.
 Note that $r^{\text{b}}_{m^{t_1+t_2}} \leq r^{\text{b}}_{m^t_1} + r^{\text{b}}_{m^{t_2}}$, i.e., the sequence $\{r^\text{b}_{m^t}\}_{t=1}^\infty$ is subadditive.\footnote{To see this, we can always concatenate the index codes for the index-coding instances with message alphabet sizes $m^{t_1}$ and $m^{t_2}$ to get an index code for the instance with message alphabet size $m^{t_1+t_2}$.} By Fekete's Subadditive Lemma,
\begin{subequations}
\begin{align}
\lim_{t \rightarrow \infty} r_{m^t}(G) &= \frac{1}{\log_2 m} \lim_{t \rightarrow \infty} \frac{r^\text{b}_{m^t}(G)}{t}\\
& = \frac{1}{\log_2 m}  \inf_{t} \frac{r^\text{b}_{m^t}(G)}{t}\\
& = \inf_{t} r_{m^t}(G), \label{eq:subadditivity}
\end{align}
\end{subequations}
for any fixed $G$ and $m$. This proves the first equality in \eqref{eq:optimal-broadcast-rate-2}. 

From definition \eqref{eq:optimal-broadcast-rate}, for any $\epsilon > 0$, we can always find some $m'$ and $t'$ such that $r_{{(m')}^{t'}}(G)< r(G) + \epsilon$, using some index code $(\phi,\{\psi_i\})$ with codewords on $\mathcal{Y}^p$. This means $\frac{p \log_2 |\mathcal{Y}|}{t' \log_2m'}< r(G) + \epsilon$. 
By concatenating this index code $b \in \mathbb{Z}^+$ times, we get codewords on $\mathcal{Y}^{bp}$ for the message alphabet size $(m')^{bt'}$, with a normalised length of $\frac{bp \log_2 |\mathcal{Y}|}{bt' \log_2m'}< r(G) + \epsilon$. Note that this concatenated code can be used as an index code for any message alphabet of size $|\mathcal{X}| = m^t$ as long as $m^t < (m')^{bt'}$ with zero padding, giving an index code of normalised length $\frac{bp \log_2 |\mathcal{Y}|}{t \log_2 m}$.
For any fixed $m', t', p,$ and $|\mathcal{Y}|$, we can choose \textbf{any} $m$ and sufficiently large integers $t$ and $b$, such that $\frac{p \log_2 |\mathcal{Y}|}{\frac{t}{b} \log_2 m} - \frac{p \log_2 |\mathcal{Y}|}{t' \log_2m'} \triangleq \eta > 0$ can be made as small as desired. Noting that $r(G) \leq r_m^t(G)$ by definition, and that $r_{m^{t}}(G) \leq \frac{bp \log_2 |\mathcal{Y}|}{t \log_2 m} < r(G) + \epsilon + \eta$ for any arbitrarily small $\epsilon, \eta >0$, we have the second equality in \eqref{eq:optimal-broadcast-rate-2} for any $m$.
\end{IEEEproof}





It follows from Proposition~\ref{prop:equivalent-beta} and subadditivity of the sequence $\{r^\text{b}_{m^t}\}_{t=1}^\infty$ that, for any $m$ and $t$,
\begin{equation}
r(G) \leq r_{m^t}(G) \leq r_{m^1}(G).
\end{equation}

We say that
$r(G)$ is the (normalised) optimal \textit{asymptotic} index codelength for $G$, when the length of the message vector, $t$, tends to infinity, and
$r_{m^t}(G)$ is the optimal \textit{finite-length} index codelength, where the messages are each a length-$t$ vector over an alphabet of size $m$. The latter is also known as the \textit{one-shot} index codelength~\cite{tahmasbishahrasbigohari15}.

\begin{remark}
We will see in Section~\ref{sec:binary} later that choosing $|\mathcal{Y}| = m$ for the finite-length case, i.e., finite $t$, may give a suboptimal index codelength.
\end{remark}

\subsection{Linear codes}

\begin{definition}
(Linear codes) Re-write the encoding function as $\phi = [\phi_1 \phi_2 \dotsm \phi_p]$, where $\phi_i: \mathcal{X}^n \mapsto \mathcal{Y}$, and consider the following three cases:
\begin{enumerate}
\item $\mathcal{X}= \mathcal{Y}= \mathbb{F}_q$, where $\mathbb{F}_q$ is a $q$-element finite field for some prime power~$q$: If each $\phi_i$ is a linear function over the field $\mathbb{F}_q$, i.e., $\phi_i(\bm{X}) = \sum_{j=1}^n k_{ij}X_j \in \mathbb{F}_q$, for some $k_{ij} \in \mathbb{F}_q$, the index code is \textit{scalar linear over the field} $\mathbb{F}_q$.
\item $\mathcal{X} = \mathbb{F}_q^t$ and $\mathcal{Y}=\mathbb{F}_q$: If $\phi_i: \mathbb{F}_q^{t \times n} \mapsto \mathbb{F}_q$ is a linear function over $\mathbb{F}_q$, then the index code is \textit{vector linear over the field} $\mathbb{F}_q$.
\item $\mathcal{X} = \mathcal{Y} = \mathcal{F}$, for any finite alphabet $\mathcal{F}$: Without loss of generality, let $\mathcal{F} = \{0,1,\dotsc, |\mathcal{F}|-1\}$. If $\phi_i: \mathcal{F}^n \mapsto \mathcal{F}$ is linear, meaning that $\phi_i(\bm{X}) = \sum_{j=1}^n k_{ij}X_j$, where the addition and the multiplication are defined over integer modulo-$|\mathcal{F}|$, then the index code is \textit{scalar linear over the ring} $\mathcal{F}$.
\item $\mathcal{X} = \mathcal{F}^t$ and $\mathcal{Y} = \mathcal{F}$, where $\mathcal{F} = \{0,1,\dotsc, |\mathcal{F}|-1\}$: If $\phi_i: \mathcal{F}^{t \times n} \mapsto \mathcal{F}$ is a linear function over integer modulo-$|\mathcal{F}|$, then the index codes is \textit{vector linear over the ring} $\mathcal{F}$.
\end{enumerate}
\end{definition}

\section{Related Results and Main Contributions} \label{sec:main}

\subsection{Existing lower bounds}

Bar-Yossef, Birk, Jayram, and Kol~\cite{baryossefbirk11} proposed a graph-theoretic lower bound on $r_{2}(G)$, by considering its acyclic subgraph. Denote the number of vertices in a \textit{maximum acyclic induced subgraph}\footnote{It is defined as an induced subgraph with the largest number of vertices.} (MAIS) of $G$ by $\mais(G)$.
The lower bound is readily extended to any message size, including the asymptotic case, as follows:
\begin{lemma}\label{lemma:graph-lower-bound}
For any $m$ and $t$,
\begin{equation}
\mais(G) \leq r(G) \leq r_{m^t}(G). \label{eq:mais-lower-bound}
\end{equation}
\end{lemma}

Let the random variables of an index code be defined as $[Y_1 Y_2 \dotsm Y_p] = \bm{Y} = \phi(\bm{X})$.
Blasiak, Kleinberg, and Lubetzky~\cite{blasiakkleinberglubertzky13} proposed a lower bound by showing that the joint entropies of $\{\bm{X},\bm{Y}\}$ must satisfy the following constraints:
\begin{enumerate}
\item \textit{Decodability:} Consider any receiver~$i \in [n]$. Knowing $\bm{Y}$ and $\bm{X}_{\on{G}{i}}$, the receiver can decode $X_i$. This means $H(X_i,\bm{X}_{\on{G}{i}},\bm{Y}) = H(\bm{X}_{\on{G}{i}},\bm{Y})$, for each $i \in [n]$.
\item \textit{Submodularity of entropy:} For two subsets of random variables $\mathcal{S}$ and $\mathcal{T}$, we have $H(\mathcal{S}) + H(\mathcal{T}) \geq H(\mathcal{S} \cup \mathcal{T}) + H(\mathcal{S} \cap \mathcal{T})$.
\item \textit{Non-Shannon-type information inequalities:} See Zhang and Yeung~\cite{zhangyeung97} for example.
\end{enumerate}
These (in)equalities can be used to obtain lower bounds to $H(\bm{Y})$, and $H(\bm{Y})$   in turn bounds the index codelength from below as $H(\bm{Y})/\log_2|\mathcal{X}| \leq \sum_{i=1}^p H(Y_i)/\log_2|\mathcal{X}| \leq \ell_{m^t}(G)$, for any index code and any choice of $m$ and $t$. 

There are infinitely many non-Shannon inequalities, and invoking all of them leads to $r(G)$~\cite[page~37]{blasiak13}.

Noting that $r_{m^t}(G') \leq r_{m^t}(G)$, for any vertex-induced subgraph $G'$ of $G$~\cite[Proposition~9]{baryossefbirk11}, the MAIS lower bound~\eqref{eq:mais-lower-bound} can also be obtained from the decodability constraints.

In this paper, we will construct MAISs and inequalities invoking the first two types of constraints (i.e., decodability and submodularity) to obtain lower bounds on $r_{m^t}(G)$.
Non-Shannon-type information inequalities are not required for the analyses in this paper.

\subsection{Existing upper bounds (achievability)}

By choosing $\mathcal{Y} = \mathcal{X}$, and sending the messages uncoded, we get an index code of length $n=|V(G)|$. This gives the following trivial upper bound on the optimal index codelength:
\begin{equation}
r(G) \leq r_{m^t}(G) \leq n.
\end{equation}

Consider the special case where each message is a binary bit, i.e., $\mathcal{X} = \mathbb{F}_2$. A scalar linear code can be formed by solving a graph function \textit{minrank}. Consider a matrix $A$ with binary elements. We say that a binary $n$-by-$n$ matrix $M$ \textit{fits} $G$ if
\begin{equation}
m_{i,j} =
\begin{cases}
1, & \text{if } i=j\\
0, & \text{if } (i \rightarrow j) \notin A(G),
\end{cases}
\end{equation}
where $m_{i,j}$ is the element of $M$ in the $i$-th row and $j$-th column.
The rest of the elements can be either 0 or 1. Denote the rank of $M$ over $\mathbb{F}_2$ by $\mathsf{rk}_2(M)$, the minrank of $M$ over $\mathbb{F}_2$ is defined as
\begin{equation}
\minrank(G) \triangleq \min \{ \mathsf{rk}_2(M): M \text{ fits } G\}.
\end{equation}

Bar-Yossef el al.~\cite{baryossefbirk11} proved the following lemma:
\begin{lemma}
\begin{equation}
r_{2}(G) \leq \minrank(G).
\end{equation}
Furthermore, if we restrict the encoding function $\phi(\bm{X})$ to be scalar linear, then
\begin{equation}
\min_{\phi,\{\psi_i\}:\phi \text{ is scalar linear}} r_2(G) = \minrank(G). 
\end{equation}
\end{lemma}

Blasiak et al.\ extended $\mathsf{minrk}_2$ to higher field sizes, $\mathcal{X} = \mathbb{F}_q^t$, obtained a similar upper bound, and showed that the bound is tight if the encoding function is restricted to be scalar or vector linear.

 Both the MAIS upper bound and the minrank lower bound are NP-hard to compute~\cite{karp72,peeters96}, and both have been shown to be loose in some instances~\cite{baryossefbirk11,lubertzkystav09}. This implies that linear or vector-linear index codes, though having practical advantages of simplifying encoding and decoding, are not necessarily optimal.

Other upper bounds can be obtained by finding disjoint cycles~\cite{neelytehranizhang13}, disjoint cliques~\cite{birkkol2006}, a special structure called \textit{interlinked cycles}~\cite{thapaongjohnson17it}, the local chromatic number of the graph~\cite{shanmugamdimakislangberg13}, or the maximum out-degree of graph partitions~\cite{birkkol2006}. Some of these approaches require $\mathcal{X}$ to be a finite field of a sufficiently large size.

Some approaches use information-theoretic random coding~\cite{arbabjolfaei13} and techniques for rate distortion~\cite{unalwagner16}. As expected, these approaches are non-constructive, and the results are asymptotic in the limit when the message alphabet size $|\mathcal{X}|$ tends to infinity.\footnote{Unlike this paper, a broadcast rate in those works~\cite{arbabjolfaei13,unalwagner16} is said to be achievable if the decoding error tends to zero as the message alphabet size tends to infinity. However, Langberg and Effros~\cite{langbergeffros11} showed that any index-coding broadcast rate (or codelength) that is achievable in the diminishing-error sense is also achievable in the zero-error sense.} Consequently, these results are upper bounds on $r(G)$, and not on $r_{m^t}(G)$ for any finite $m$ and $t$.

We will show that, for most cases considered in this paper, the interlinked-cycle cover can be used to obtain optimal scalar linear codes. Here, we briefly describe the scheme:
\begin{definition} \label{definition:interlinked-cycles}
(Interlinked cycle~\cite{thapaongjohnson17it}) A directed subgraph $G$ is an interlinked cycle if and only if we can find a vertex subset $V_\text{I} \subseteq V(G)$, called an inner-vertex set, such that
\begin{enumerate}
\item there is no directed cycle in $G$ that contains one and only one inner vertex, and
\item for any ordered pair of inner vertices $(i,j)$, there is one and only one path from $i$ to $j$, in which all other vertices in the path, if exists, are not in $V_\text{I}$.
\end{enumerate}
\end{definition}

\begin{definition} \label{definition:ic-cover}
  (Interlinked-cycle cover~\cite{thapaongjohnson17it}) Given an interlinked cycle~$G$ with an inner-vertex set $V_\text{I}$ and a message alphabet $\mathcal{X}$, a scalar linear code of length $|V(G)| - |V_\text{I}| + 1$ over a ring with $|\mathcal{X}|$ elements can formed as follows:
\begin{align}
&\sum_{i \in V_\text{I}} X_i, \\
&X_j + \sum_{k \in \on{G}{j}} X_k, \quad \text{for each } j \in V(G) \setminus V_\text{I}.
\end{align}
\end{definition}

Note that a cycle $C$ of length $L$, for any $L \geq 2$, with vertices and arcs $c_1 \rightarrow c_2 \rightarrow \dotsm \rightarrow c_L \rightarrow c_1$ is a special case of interlinked cycles, by choosing any two vertices therein to be the inner-vertex set. For example, by choosing $\{c_{L-1},c_L\}$ to be the inner-vertex set, we have the following scalar linear index code of length $|V(C)|-1$ for $C$:
\begin{equation}
X_1 + X_2,\;  X_2 + X_3, \; \dotsc, \; X_{L-1} + X_L. \label{eq:cycle}
\end{equation}
The above code~\eqref{eq:cycle}, also known as a cyclic code, was used by Neely, Tehrani, and Zhang~\cite{neelytehranizhang13} and Ong, Ho, and Lim~\cite{ongholim16}.

Also note that a clique $Q$ (a subgraph in which each vertex has an outgoing arc to every other vertex) is an interlinked cycle with all its vertices as the inner-vertex set. The interlinked-cycle cover gives an index code of length 1: $\sum_{i \in V(Q)} X_i$. This is also known as the clique cover~\cite{baryossefbirk11}.

Recall that $N_G^+(i)$ is the out-neighbourhood of $i$ in $G$. Let $N_G^-(i)$ denote the in-neighbourhood of $i$ in $G$.

\begin{definition} \label{def:super-vertex}
  (Interlinked cycle with super vertices~\cite{thapaongjohnson17it}) Consider a vertex set $V_\text{s}$ in a graph $G$ satisfying the following conditions: For all distinct pairs $i,j \in V_\text{s}$, we have
  \begin{itemize}
    \item $(i \rightarrow j) \in A(G)$, i.e., all vertices in $V_\text{s}$ have arcs to each other, and
    \item $N_G^+(i) \setminus V_\text{s} = N_G^+(j) \setminus V_\text{s}$ and $N_G^-(i) \setminus V_\text{s} = N_G^-(j) \setminus V_\text{s}$, i.e., all vertices in $V_\text{s}$ have the same incoming and outgoing connection to vertices outside $V_\text{s}$ in $G$.
  \end{itemize}
We can define a new graph $G'$ by replacing $V_\text{s}$ (and all arcs to and from these vertices) by a \textit{super vertex}, say $p$, with $N_{G'}^+(p) = N_G^+(i) \setminus V_\text{s}$ and $N_{G'}^-(p) = N_G^-(i) \setminus V_\text{s}$, for any arbitrarily chosen $i \in V_s$. If $G'$ is an interlinked cycle with an inner-vertex set $V_\text{I}$, where $p \notin V_I$, then we say that $G$ is an interlinked cycle with an inner-vertex set $V_\text{I}$ and a super-vertex set $V_\text{s}$. The index code formed by the interlinked-cycle cover for $G'$ is an index code (of the same length) for $G$ with $X_p$ replaced by $\sum_{i \in V_\text{s}}X_i$.
\end{definition}




\subsection{Existing capacity results}

Although there are several different approaches to computing upper bounds on $r(G)$ and $r_{m^t}(G)$, it is not easy to determine when these bounds are tight (or not). We now present a few classes of graphs where the bounds have been shown to be tight.

Bar-Yossef et al.~\cite{baryossefbirk11} showed that if $G$ is acyclic, then $r(G) = r_{m^t}(G) = |G| = \mais(G)$.

Consider a special class of graphs $G$ where $(i \rightarrow j) \in A(G)$ if and only if $(j \rightarrow i) \in A(G)$. This models index-coding instances with symmetrical knowledge, i.e., if receiver~$i$ knows $x_j$, then receiver~$j$ knows $x_i$. Any graph $G$ of this type can be mapped to a corresponding undirected graph $G_\text{u}$ with the same vertex set as $G$, and an edge $(i,j) \in E(G_\text{u})$ exists if and only if $(i \rightarrow j) \in A(G)$. Bar-Yossef et al.~\cite{baryossefbirk11} found $r_{2}(G_\text{u})$ for the following classes of undirected side-information graphs:
\begin{itemize}
\item $G_\text{u}$ is a perfect graph,
\item $G_\text{u}$ is an odd hole where $|V(G)| \geq 5$, or
\item $G_\text{u}$ is an odd anti-hole where $|V(G)| \geq 5$.
\end{itemize}

Neely et al.~\cite{neelytehranizhang13} showed that if $G$ consists of disjoint cycles, then $r(G) = r_{m^t}(G) = |V(G)| - N_\text{cycle}$ for all $m$ and $t$, where $N_\text{cycles}$ is the number of cycles (all being disjoint) in $G$. This is commonly known as \textit{cycle cover}. Yu and Neely~\cite{yuneely14} represented index-coding instances using bipartite graphs, and found $r(G)$  for all planar bipartite graphs. 

It has been verified by intensive computer calculations that composite coding~\cite{arbabjolfaei13} (derived using random-coding arguments) is optimal for all $G$ with $|V(G)| \leq 5$, giving $r(G)$.

Unal and Wagner~\cite{unalwagner16} derived the asymptotic optimal index codelength for all general (in the sense that each message can be requested by several receivers) index-coding instances up to three receivers. Their method is based on rate-distortion theory, which also uses random-coding arguments.

To find $r_{m^t}(G)$ by brute force, one can form the \textit{confusion graph}~\cite{baryossefbirk11} of $G$ with $m^{tn}$ vertices, and calculate the chromatic number (which is an NP-complete problem) of the confusion graph. This method is, however, intractable as the order of the confusion graph grows exponentially with $tn$.

\subsection{Main results of this paper}

The main results of this paper are as follows: We find the optimal index codelength and the minimum message alphabet size required to achieve the optimal index codelength  for the following classes of index-coding instances:
\begin{enumerate}
\item (In Section~\ref{section:result-1}) For any $G$ that can be made acyclic after removing two or fewer arcs: We derive $r(G)$, and show that $r(G) = r_{m^t}(G)$ for all integers $m \geq 2$ and $t \geq 1$.
\item (In Section~\ref{sec:result-2}) For any $G$ of up to five vertices (there are 9847 non-isomorphic graphs in total):
\begin{enumerate}
\item For 9819 non-isomorphic graphs, we derive $r(G)$, and show that $r(G) = r_{m^t}(G)$ for all integers $m \geq 2$ and $t \geq 1$.
\item For the remaining 28 non-isomorphic graphs, we derive $r_{m^{2k}}(G)$, and show that $r(G) = r_{m^{2k}}(G)$, for all integers $m \geq 2$ and $k \geq 1$.
\end{enumerate}
\end{enumerate}

Furthermore, for all the above cases, we show that linear index codes (over a ring) are optimal.

Recall that $r_{2}(G)$ is the solution for an index-coding instance $G$ where each message consists of a single binary bit. The above result of $r(G) = r_{2}(G)$, together with linear codes in $\mathbb{F}_2$ being optimal, means that encoding can be done bit-by-bit without loss of optimality. The advantages of this are that (i) encoding is simple (bit-wise XOR of the messages), and that (ii) decoding is instantaneous. For cases where $r(G) = r_{2^2}(G)$, we can achieve the optimal broadcast rate by encoding (and decoding) two bits of messages at a time.

\section{Optimal Index Codelength when $\mais(G) \geq |V(G)| - 2$} \label{section:result-1}
\subsection{Main result}
In this section, we show the following theorem:
\begin{theorem} \label{thm:mais}
If $\mais(G) \geq |V(G)| - 2$, then 
\begin{equation}
r(G) = r_{m^t}(G) = \minrank(G) = \mais(G), \label{eq:capacity-mais}
\end{equation}
for any integers $m\geq 2$ and $t \geq 1$, 
and the minimum index codelength is achievable using scalar linear codes over a ring with $m^t$ elements.
\end{theorem}

It follows from Theorem~\ref{thm:mais} that the minimum alphabet size required to achieve $r(G)$ is $|\mathcal{X}|=2$, i.e., binary messages.

This theorem will be used to establish the result for all graphs up to five vertices in Section~\ref{sec:result-2}.

\begin{remark}
Characterising graphs having a certain $\minrank(G)$ value is hard. Dau et al.~\cite{dauskachekcheeisit12} managed to  characterise all \textit{undirected} graphs whose $\minrank(G_\text{u})$ is $|V(G_\text{u})|-2$ or $|V(G_\text{u})|-1$, and all \textit{directed} graphs whose $\minrank(G)$ is 2 or $|V(G)|$. They are, however, unable to characterise directed graphs whose $\minrank(G)$ is $|V(G)|-1$ or $|V(G)|-2$. 
For any directed graph $G$ whose $\mais(G)$ equals $|V(G)|-1$ or $|V(G)|-2$, we show in this paper that linear index codes are optimal, meaning that $\mais(G)=\minrank(G)$. So, we have incidentally characterised a subset of directed graphs whose $\minrank(G)$ equals $|V(G)|-1$ or $|V(G)|-2$.
\end{remark}





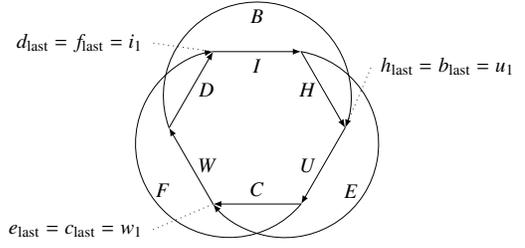
\begin{figure}[t]
\centering
\resizebox{70mm}{!}{%
\begin{tikzpicture}
\def\R{15mm}

\foreach \x in {1,2,...,6}{
\coordinate (\x) at ({60*(\x-1)}:\R);
}

\draw[->,>=latex] (1) -- node[midway,left] {$U$} (6);
\draw[->,>=latex] (6) -- node[midway,above] {$C$} (5);
\draw[->,>=latex] (5) -- node[midway,right] {$W$} (4);
\draw[->,>=latex] (4) -- node[midway,right] {$D$} (3);
\draw[->,>=latex] (3) -- node[midway,below] {$I$} (2);
\draw[->,>=latex] (2) -- node[midway,left] {$H$} (1);

\draw[->,>=latex] (4) arc (200:-20:1.065*\R) node[midway,below] {$B$} ;
\draw[->,>=latex] (2) arc (80:-140:1.065*\R) node[midway,left] {$E$} ;
\draw[->,>=latex] (6) arc (320:100:1.065*\R) node[midway,right] {$F$} ;

\draw[dotted] (1) -- +({0.3*\R},{+0.7*\R}) node[inner sep=0,label=right:{$h_\text{last}=b_\text{last} = u_1$}] {};
\draw[dotted] (5) -- +({-0.7*\R},{-0.3*\R}) node[inner sep=0,label=left:{$e_\text{last} = c_\text{last}=w_1$}] {};
\draw[dotted] (3) -- +({-0.7*\R},{+0.1*\R}) node[inner sep=0,label=left:{$d_\text{last}=f_\text{last} = i_1$}] {};

\end{tikzpicture}
}%
\caption{An important element in proving Theorem~\ref{thm:mais} is to show that if $\mais(G) = |V(G)|-2$ and condition (3.ii) is true, then $G$ must contain a subgraph $G'$ shown above. Here, every arrow represents a path, which is denoted by a capital letter. The paths do not share common vertices except the end points.
Vertices in each path is denoted by the corresponding small letter, indexed in the direction of the arcs, e.g., path $C$ is $c_1 \rightarrow c_2 \rightarrow \dotsm \rightarrow c_\text{last}$.
All paths except $I$, $W$, and $U$ must contain one or more arcs.}
\label{fig:three-cycles}
\end{figure}

\begin{IEEEproof}[Proof of Theorem~\ref{thm:mais}]
It follows from Lemma~\ref{lemma:graph-lower-bound} that $\mais(G)$ is a lower bound on $r_{m^t}(G)$. So, we only need to prove achievability. 


Without loss of generality, let $\mathcal{X} = \{0,1,\dotsc,|\mathcal{X}|-1\}$.
We will show that scalar linear codes over the ring $\mathcal{X}$ is optimal. To this end, we choose $\mathcal{Y} = \mathcal{X}^p$, and therefore the normalised codelength is given by $\ell_{m^t}(G) = p$.

\subsubsection{$\mais(G) = |V(G)|$} For this case, $G$ is acyclic. As mentioned in the previous section, sending all messages uncoded (i.e., $\phi(\bm{X}) = \bm{X}$, and hence we have a linear code of length $\ell_{m^t}(G) = p = |V(G)| = n$) achieves the MAIS lower bound, and we have \eqref{eq:capacity-mais}.

\subsubsection{$\mais(G) = |V(G)|-1$} For this case, the directed graph $G$ must contain at least one cycle; otherwise, $\mais(G) = |V(G)|$. Let the cycle be $C \subseteq G$.

 We send a cyclic code for $C$ and the rest of the messages $\bm{X}_{V(G) \setminus V(C)}$ uncoded, forming an index code of length $|V(G)|-1$. The cyclic code allows all receivers $i \in V(C)$ to decode $X_i$. In addition, all receivers $j \in V(G) \setminus V(C)$ can decode $X_j$ as the messages are sent uncoded.

\subsubsection{$\mais(G) = |V(G)| - 2$} There are two possibilities for $G$:
\begin{enumerate}
\item[](3.i) There are two vertex-disjoint cycles, or 
\item[](3.ii) There are no two vertex-disjoint cycles.
\end{enumerate}
For case (3.i), we code each of the two disjoint cycles with a cyclic code, and send the rest of the messages in $G$ uncoded. This achieves a codelength of $|V(G)|-2$.\footnote{Let the two disjoint cycles be $C_1$ and $C_2$. The two cyclic codes, each for one cycle, are of length $|V(C_1)|-1$ and $|V(C_2)|-1$ respectively. Together with uncoded messages with a total length $|V(G)|-|V(C_1)|-|V(C_2)|$, we get an overall index codelength of $|V(G)|-2$.}

For case (3.ii), we will derive Lemmas~\ref{lemma:structure} (stated next), which says that if $\mais(G) = |V(G)|-2$ and there are no two vertex-disjoint cycles, then $G$ contains a subgraph $G'$ of the form depicted in Figure~\ref{fig:three-cycles}, in which each arrow represents a path.

Now, note that $G'$ is an interlinked cycle with inner-vertex set $\{i_1, u_1, w_1\}$. Here, for path $U$, we label the vertices in the path as $u_1 \rightarrow u_2 \rightarrow \dotsm u_\text{last}$. Using the interlinked-cycle cover, we obtain a scalar linear code of length $|V(G')|-2$ over the ring $\mathcal{X}$ for $G'$. Combining this with sending the remaining messages $\bm{X}_{V(G \setminus G')}$ uncoded gives an index code with a length of $|V(G)|-2$. 
\end{IEEEproof}

\begin{remark}
In a conference version of this paper~\cite{ongnetcod14}, we presented an alternative coding scheme that constructs another scalar linear code of length $|V(G')|-2$ for $G'$.
\end{remark}

\subsection{Existence of a special structure: Figure~\ref{fig:three-cycles}}

It is easy to obtain a \textit{saving} (which is the reduction of codelength compared to sending uncoded messages) of one for each vertex-disjoint cycle using a simple cyclic code. The main challenge of Theorem~\ref{thm:mais} is to show that for case (3.ii), even though we cannot find two vertex-disjoint cycles, we can achieve a saving of two. The following lemma is a key step.

\begin{lemma} \label{lemma:structure}
If $\mais(G) = |V(G)| - 2$ , and there are no two vertex-disjoint cycles, then $G$ must contain a subgraph (not necessarily an induced subgraph) shown in Figure~\ref{fig:three-cycles}.
\end{lemma}

\begin{IEEEproof} 
See Appendix~\ref{appendix:structure}.
\end{IEEEproof}

\section{Optimal Index Codelength for All Graphs up to Five Vertices}
\label{sec:result-2}

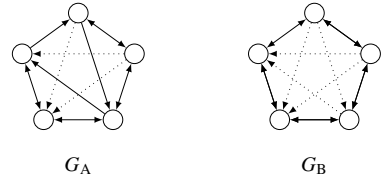
\begin{figure}[t]
\centering
\resizebox{50mm}{!}{%
\begin{tikzpicture}
\def\D{12mm}

\graph [nodes={draw, circle}, empty nodes, clockwise, n=5] {
{[path,<->, edges={>=latex}] 1, 2, 3, 4, 5},
{[edges={>=latex}] 5 -> 1 -> 3 -> 5}, 
{[edges={dotted,>=latex}] 1 -> 4,  2 -> 4,  2 -> 5}
};
\node[label={[align=center]below:$G_\text{A}$}] at (0,{-\D}) {};

\begin{scope}[xshift=40mm]
\graph [nodes={draw, circle}, empty nodes, clockwise, n=5] {
1,2,3,4,5,
{[path,<->, edges={>=latex}] 1  <-> 2  <-> 3  <-> 4  <-> 5  <-> 1},
{[edges={dotted,>=latex}] 1 -> 3 -> 5 , 2 -> 5 , 2 -> 4 , 1 -> 4 }
};
\node[label={[align=center]below:$G_\text{B}$}] at (0,{-\D}) {};
\end{scope}

\end{tikzpicture}
}%

\caption{Let $\mathcal{G}_\text{s}$ be a set of 28 non-isomorphic five-vertex graphs, formed by by removing any number (zero to three inclusive) of dotted arcs from $G_\text{A}$ (this gives eight non-isomorphic graphs) and removing any number of dotted arcs from $G_\text{B}$ (this gives 20 non-isomorphic graphs).}
\label{fig:special-group}
\end{figure}

\ifx\doublecolumn\undefined
\begin{table*}
\centering
\begin{tabular}{|c|l !{\VRule[2pt]} c|c|c|c|c|c|c|c|c|c|}
\hline
\multicolumn{2}{|c!{\VRule[2pt]}}{Number of receivers, $|V(G)|$} & 1 & 2 & \multicolumn{2}{c|}{3} & \multicolumn{2}{c|}{4} & \multicolumn{4}{c|}{5} \\
\hline
\multicolumn{2}{|c!{\VRule[2pt]}}{\multirow{3}{*}{Number of non-isomorphic $G$}} & 1 & 3 & \multicolumn{2}{c|}{16} & \multicolumn{2}{c|}{218} & \multicolumn{4}{c|}{9608} \\
\cline{3-12}
\multicolumn{2}{|c!{\VRule[2pt]}}{} & 1 & 3 & 9 & 7 & 41 & \parbox{10ex}{\centering 177} & \parbox{10ex}{\centering 334} & \multicolumn{2}{c|}{($\mathcal{G}_\text{s}$)} & \parbox{25ex}{\centering 9246} \\
\multicolumn{2}{|c!{\VRule[2pt]}}{} &  & & & & & &  & 1 & 27  & \\
\specialrule{2pt}{0pt}{0pt}
\multicolumn{2}{|c!{\VRule[2pt]}}{Binary messages, i.e., $m=2$, $t=1$} & \cellcolor{yellow} & \cellcolor{yellow} & \multicolumn{1}{c}{\cellcolor{yellow}} & \cellcolor{green} & \multicolumn{1}{c}{\cellcolor{yellow}} & \cellcolor{green} & \multicolumn{1}{c}{\cellcolor{yellow}} & \multicolumn{1}{c}{\cellcolor{yellow} $\ast$} & \multicolumn{1}{c}{} &  \multicolumn{1}{c|}{\cellcolor{green}}\\
\hline
\multirow{9}{*}{Messages of size $m^t$} & $\forall m \geq 3,t=1$ & \cellcolor{yellow} & \cellcolor{yellow} & \multicolumn{1}{c}{\cellcolor{yellow}} & \cellcolor{green} & \multicolumn{1}{c}{\cellcolor{yellow}} & \cellcolor{green} & \multicolumn{1}{c}{\cellcolor{yellow}} & \multicolumn{2}{c}{(28)} & \cellcolor{green}\\
 & $\forall m \geq 2,t=2$ & \cellcolor{yellow} & \cellcolor{yellow} & \multicolumn{1}{c}{\cellcolor{yellow}} & \cellcolor{green} & \multicolumn{1}{c}{\cellcolor{yellow}} & \cellcolor{green} & \multicolumn{1}{c}{\cellcolor{yellow}} & \multicolumn{3}{c|}{\cellcolor{green}}\\
 & $\vdots$ & \cellcolor{yellow} & \cellcolor{yellow} & \multicolumn{1}{c}{\cellcolor{yellow}} & \cellcolor{green} & \multicolumn{1}{c}{\cellcolor{yellow}} & \cellcolor{green} & \multicolumn{1}{c}{\cellcolor{yellow}} & \multicolumn{2}{c}{$\vdots$} & \cellcolor{green}\\
 &$\forall m \geq 2$, odd $t$ & \cellcolor{yellow} & \cellcolor{yellow} & \multicolumn{1}{c}{\cellcolor{yellow}} & \cellcolor{green} & \multicolumn{1}{c}{\cellcolor{yellow}} & \cellcolor{green} & \multicolumn{1}{c}{\cellcolor{yellow}} & \multicolumn{2}{c}{(28)} & \cellcolor{green}\\
 &$\forall m \geq 2$, even $t$ & \cellcolor{yellow} & \cellcolor{yellow} & \multicolumn{1}{c}{\cellcolor{yellow}} & \cellcolor{green} & \multicolumn{1}{c}{\cellcolor{yellow}} & \cellcolor{green} & \multicolumn{1}{c}{\cellcolor{yellow}} & \multicolumn{3}{c|}{\cellcolor{green}}\\
 & $\vdots$ & \cellcolor{yellow} & \cellcolor{yellow} & \multicolumn{1}{c}{\cellcolor{yellow}} & \cellcolor{green} & \multicolumn{1}{c}{\cellcolor{yellow}} & \cellcolor{green} & \multicolumn{1}{c}{\cellcolor{yellow}} & \multicolumn{2}{c}{$\vdots$} & \cellcolor{green}\\
 & $\forall m \geq 2,t=\infty$ & \multicolumn{10}{c|}{\cellcolor{pink}}\\
\hline
\multicolumn{12}{l}{Note: The column width is not indicative of the number of non-isomorphic graphs.}\\
\multicolumn{12}{l}{Legend:\quad \colorbox{yellow}{\color{yellow}$\ast$} \colorbox{yellow}{$\ast$} Solved by Bar-Yossef et al.~\cite{baryossefbirk11}\quad\quad \colorbox{pink}{\color{pink} $\ast$} Solved by Arbabjolfaei et al.\cite{arbabjolfaei13} \quad\quad\colorbox{yellow}{\color{yellow} $\ast$} \colorbox{pink}{\color{pink} $\ast$} \colorbox{green}{\color{green} $\ast$} Solved in this paper}\\
\multicolumn{12}{l}{}
\end{tabular}
\caption{Graphs for which the optimal index codelength is found for message of size $m^t$}
\label{table:1}
\end{table*}
\else

\fi

\ifx\doublecolumn\undefined
\else

\begin{table*}
\centering
\begin{tabular}{|c|l !{\VRule[2pt]} c|c|c|c|c|c|c|c|c|c|}
\hline
\multicolumn{2}{|c!{\VRule[2pt]}}{Number of receivers, $|V(G)|$} & 1 & 2 & \multicolumn{2}{c|}{3} & \multicolumn{2}{c|}{4} & \multicolumn{4}{c|}{5} \\
\hline
\multicolumn{2}{|c!{\VRule[2pt]}}{\multirow{3}{*}{Number of non-isomorphic $G$}} & 1 & 3 & \multicolumn{2}{c|}{16} & \multicolumn{2}{c|}{218} & \multicolumn{4}{c|}{9608} \\
\cline{3-12}
\multicolumn{2}{|c!{\VRule[2pt]}}{} & 1 & 3 & 9 & 7 & 41 & \parbox{10ex}{\centering 177} & \parbox{10ex}{\centering 334} & \multicolumn{2}{c|}{($\mathcal{G}_\text{s}$)} & \parbox{25ex}{\centering 9246} \\
\multicolumn{2}{|c!{\VRule[2pt]}}{} &  & & & & & &  & 1 & 27  & \\
\specialrule{2pt}{0pt}{0pt}
\multicolumn{2}{|c!{\VRule[2pt]}}{Binary messages, i.e., $m=2$, $t=1$} & \cellcolor{lightgray} & \cellcolor{lightgray} & \multicolumn{1}{c}{\cellcolor{lightgray}} & \cellcolor{black} & \multicolumn{1}{c}{\cellcolor{lightgray}} & \cellcolor{black} & \multicolumn{1}{c}{\cellcolor{lightgray}} & \multicolumn{1}{c}{\cellcolor{lightgray} $\ast$} & \multicolumn{1}{c}{} &  \multicolumn{1}{c|}{\cellcolor{black}}\\
\hline
\multirow{9}{*}{Messages of size $m^t$} & $\forall m \geq 3,t=1$ & \cellcolor{lightgray} & \cellcolor{lightgray} & \multicolumn{1}{c}{\cellcolor{lightgray}} & \cellcolor{black} & \multicolumn{1}{c}{\cellcolor{lightgray}} & \cellcolor{black} & \multicolumn{1}{c}{\cellcolor{lightgray}} & \multicolumn{2}{c}{(28)} & \cellcolor{black}\\
 & $\forall m \geq 2,t=2$ & \cellcolor{lightgray} & \cellcolor{lightgray} & \multicolumn{1}{c}{\cellcolor{lightgray}} & \cellcolor{black} & \multicolumn{1}{c}{\cellcolor{lightgray}} & \cellcolor{black} & \multicolumn{1}{c}{\cellcolor{lightgray}} & \multicolumn{3}{c|}{\cellcolor{black}}\\
 & $\vdots$ & \cellcolor{lightgray} & \cellcolor{lightgray} & \multicolumn{1}{c}{\cellcolor{lightgray}} & \cellcolor{black} & \multicolumn{1}{c}{\cellcolor{lightgray}} & \cellcolor{black} & \multicolumn{1}{c}{\cellcolor{lightgray}} & \multicolumn{2}{c}{$\vdots$} & \cellcolor{black}\\
 &$\forall m \geq 2$, odd $t$ & \cellcolor{lightgray} & \cellcolor{lightgray} & \multicolumn{1}{c}{\cellcolor{lightgray}} & \cellcolor{black} & \multicolumn{1}{c}{\cellcolor{lightgray}} & \cellcolor{black} & \multicolumn{1}{c}{\cellcolor{lightgray}} & \multicolumn{2}{c}{(28)} & \cellcolor{black}\\
 &$\forall m \geq 2$, even $t$ & \cellcolor{lightgray} & \cellcolor{lightgray} & \multicolumn{1}{c}{\cellcolor{lightgray}} & \cellcolor{black} & \multicolumn{1}{c}{\cellcolor{lightgray}} & \cellcolor{black} & \multicolumn{1}{c}{\cellcolor{lightgray}} & \multicolumn{3}{c|}{\cellcolor{black}}\\
 & $\vdots$ & \cellcolor{lightgray} & \cellcolor{lightgray} & \multicolumn{1}{c}{\cellcolor{lightgray}} & \cellcolor{black} & \multicolumn{1}{c}{\cellcolor{lightgray}} & \cellcolor{black} & \multicolumn{1}{c}{\cellcolor{lightgray}} & \multicolumn{2}{c}{$\vdots$} & \cellcolor{black}\\
 & $\forall m \geq 2,t=\infty$ & \multicolumn{10}{c|}{\cellcolor{gray}}\\
\hline
\multicolumn{12}{l}{Note: The column width is not indicative of the number of non-isomorphic graphs.}\\
\multicolumn{12}{l}{Legend:\quad \colorbox{lightgray}{\color{lightgray}$\ast$} \colorbox{lightgray}{$\ast$} Solved by Bar-Yossef et al.~\cite{baryossefbirk11}\quad\quad \colorbox{gray}{\color{gray} $\ast$} Solved by Arbabjolfaei et al.\cite{arbabjolfaei13} \quad\quad\colorbox{lightgray}{\color{lightgray} $\ast$} \colorbox{gray}{\color{gray} $\ast$} \colorbox{black}{\color{black} $\ast$}  Solved in this paper}\\
\multicolumn{12}{l}{}
\end{tabular}
\caption{Graphs for which the optimal index codelength is found for message of size $m^t$}
\label{table:1}
\end{table*}

\fi

In this section, we use Theorem~\ref{thm:mais} to obtain the optimal index codelength for graphs up to five vertices. First, we define $\mathcal{G}_\text{s}$ to be a set of 28 non-isomorphic five-vertex subgraphs of the two graphs in Figure~\ref{fig:special-group}. More specifically, $\mathcal{G}_\text{s}$ consist of
\begin{itemize}
\item all eight non-isomorphic graphs formed by removing any number (zero to three inclusive) of dotted arcs of $G_\text{A}$, and
\item all 20 non-isomorphic graphs formed by removing any number (zero to five inclusive) of dotted arcs of $G_\text{B}$.
\end{itemize}
In this paper, solid, dashed, and dotted arrows are all used to represent arcs or paths (it will be clear from context whether they are arcs or paths).

Also, let $\mathcal{G}_{1:5}$ be the set of all non-isomorphic graphs up to and including five vertices. $\mathcal{G}_{1:5}$ contains 9847 non-isomorphic graphs~\cite{sloane10}. 

We now state our main results for $\mathcal{G}_{1:5} \setminus \mathcal{G}_\text{s}$ and for $\mathcal{G}_\text{s}$.

\begin{theorem} \label{thm:5-nodes-all}
For any $G \in \mathcal{G}_{1:5} \setminus \mathcal{G}_\text{s}$,
\begin{equation}
r(G) = r_{m^t}(G) = \minrank(G) = \mais(G), \label{eq:capacity-mais-theorem}
\end{equation}
for any integers $m \geq 2$ and $t \geq 1$. The optimal index codelength is achievable using \textit{scalar} linear codes over a ring with $m^t$ elements.
\end{theorem}

It follows from Theorem~\ref{thm:5-nodes-all} that for any $G \in G_{1:5} \setminus G_\text{s}$, the minimum message alphabet size required to achieve $r(G)$ is $|\mathcal{X}|=2$.


\begin{theorem} \label{thm:5-others}
For any $G \in \mathcal{G}_\text{s}$, we have that
\begin{equation}
2 = \mais(G) < r(G)  = 2.5.
\end{equation}
In addition, if $m \geq 2$ and $t=2k$ for some integer~$k \geq 1$, then
\begin{equation}
r(G) = r_{m^{2k}}(G) =  2.5, \label{eq:capacity-mais-2}
\end{equation}
and the optimal index codelength is achievable using \textit{vector} linear codes over a ring with $m^k$ elements.
\end{theorem}

It follows from Theorem~\ref{thm:5-others} that for any $G \in  \mathcal{G}_\text{s}$, the minimum message alphabet size required to achieve $r(G)$ is $|\mathcal{X}|=4$.

\begin{IEEEproof}[Proof of Theorem~\ref{thm:5-nodes-all}]
Note that for any graph, we must have that
\begin{equation}
1 \leq \mais(G) \leq |V(G)|. \label{eq:mais-constraint}
\end{equation}
We now prove Theorem~\ref{thm:5-nodes-all} by considering graphs of different orders. 
For $|V(G)| \in \{1,2,3\}$, we have
\begin{equation}
|V(G)| - 2 \leq 1 \leq \mais(G),
\end{equation}
where the second inequality follows from \eqref{eq:mais-constraint}. Invoking Theorem~\ref{thm:mais}, we get \eqref{eq:capacity-mais}.

For $|V(G)|=4$, if $\mais(G) \in \{ 2,3,4\}$, then $|V(G)| - \mais(G) \leq 2$. We again use Theorem~\ref{thm:mais} to get \eqref{eq:capacity-mais}.
For the remaining case where $\mais(G) = 1$, 
any two-vertex induced subgraph is a cycle (i.e., there are arcs in both directions between any two vertices); otherwise $\mais(G) \geq 2$. In other words, each receiver~$i$ know all other messages $\bm{X}_{[4] \setminus \{i\}}$. So, sending a length-1 index code, $X_1 + X_2 + X_3 + X_4$, satisfies all receivers' requirements, and achieves the MAIS lower bound. So, we get \eqref{eq:capacity-mais}, where the last equality is follows by observing that scalar linear codes are optimal.

For $|V(G)|=5$, if $\mais(G) \in \{ 3,4,5\}$, then again we have \eqref{eq:capacity-mais}. Also, if $\mais(G)=1$, we can use the same argument for $|V(G)|=4$ to show that the length-1 index code $X_1 + X_2 + X_3 + X_4 + X_5$ is achievable and is hence optimal.

For all the above cases, scalar linear codes over the ring $\mathcal{X}$ are optimal, and the MAIS lower bound is tight. The proof of Theorem~\ref{thm:5-nodes-all} is complete with Lemma~\ref{proposition:last-case} below, addressing the remaining case.
\end{IEEEproof}

The main challenge in proving Theorem~\ref{thm:5-nodes-all} is to show the following:

\begin{lemma} \label{proposition:last-case}
If $|V(G)| = 5$,  $G \notin \mathcal{G}_\text{s}$, and $\mais(G) = 2$, then
\begin{equation}
r(G) = r_{m^t}(G) = \minrank(G) = \mais(G), \label{eq:capacity-mais-lemma}
\end{equation}
for any integers $m \geq 2$ and $t \geq 1$. The optimal index codelength is achievable using \textit{scalar} linear codes over a ring with $m^t$ elements.
\end{lemma}

\begin{IEEEproof}[Proof of Lemma~\ref{proposition:last-case}]
See Appendix~\ref{sec:lemma-proof}.
\end{IEEEproof}

\begin{IEEEproof}[Proof of Theorem~\ref{thm:5-others}]
See Appendix~\ref{theorem-proof}.
\end{IEEEproof}

The results of Theorems~\ref{thm:5-nodes-all} and \ref{thm:5-others} in comparison with existing results are summarised in Table~\ref{table:1}. In Table~\ref{table:1}, we consider all non-isomorphic directed graphs up to and including five vertices. The columns denote distinct non-isomorphic graphs. For example, there are 218 non-isomorphic graphs with four vertices. A cell is non-white if the optimal index codelength of the corresponding graph has been found. We have used different shades to indicate different research groups that found the optimal index codelength. The rows represent the message size, given by $m^t$. 

We show using an example of $|V(G)|=4$ how to obtain Table~\ref{table:1}.
Out of the 218 non-isomorphic graphs with four vertices, Bar-Yossef et al.\ have found the optimal index codelength for 41 of them, for all message sizes $m^t$. The 41 non-isomorphic graphs consist of the following:
\begin{itemize}
\item The empty graph,\footnote{An empty graph contains no arc.} which is both acyclic and perfect\footnote{A directed graph is prefect (in the context of this paper) if it is symmetric, and the corresponding undirected graph is a perfect graph.}.
\item 30 of them that are non-empty and acyclic~\cite{sloane10acyclic}.
\item 10 of them that are non-empty and perfect~\cite{sloane10perfect}. (Note that if a graph is not empty, it cannot be both acyclic and perfect)
\end{itemize}
All the light grey cells correspond to acyclic and/or perfect graphs, except for the graph marked with an asterisk, which corresponds to the (undirected) 5-cycle. For the 5-cycle, Bar-Yossef et al.\ showed that $r_2(G)=3$ when the messages are binary. The lower bound was found by a brute-force exhaustive search.

Also shown in the table, Arbabjolfaei et al.\ found $r(G) = \lim_{r \rightarrow \infty} r_{m^t}(G)$ for all graphs up to five vertices.

Theorems~\ref{thm:5-nodes-all} and \ref{thm:5-others} cover all non-white cells in the table, except the asterisked cell.

\subsection{Optimal codelength for \textnormal{$\mathcal{G}_\text{s}$} with binary messages via the confusion-graph technique} \label{sec:binary}

Recall that $\mathcal{G}_\text{s}$ contains all (non-strict) subgraphs of $G_\text{A}$ and $G_\text{B}$ in Figure~\ref{fig:special-group} with none or some dotted arcs removed. Although, in Theorem~\ref{thm:5-others}, we have derived the optimal codelength for all $\mathcal{G} \in \mathcal{G}_\text{s}$ when $m^t$ when $t$ is any even integer, we do not have results for odd $t$.

In this section, we discuss the optimal codelength for the members of $\mathcal{G}_\text{s}$ specifically when each message is a binary bit, i.e., when $m=2$ and $t=1$. This corresponds to the cells in the 28 columns marked $\mathcal{G}_\text{s}$ and the top row in Table~\ref{table:1}.

\subsubsection{Confusion graphs}
One can use a brute-force technique of confusion graph (see Bar-Yossef et al.~\cite{baryossefbirk11} for example) to determine the optimal codelength. We first describe the notion of confusion graphs:
\begin{definition} \label{definition:confusion-graph}
For an index coding instance $G$ and a message alphabet $\mathcal{X}$, its undirected confusion graph $G_\text{confusion}$ has $|\mathcal{X}|^{|V(G)|}$ vertices. The vertices are labelled with distinct realisations of the message tuples, i.e., $\{[x_1 x_2 \dotsm x_n] \in \mathcal{X}^n\}$, where $n = |V(G)|$. An edge exists between two vertices, say $\bm{x}$ and $\bm{x}'$, if and only if there exists a receiver~$j \in [n]$ such that
\begin{align}
&& x_j &\neq x'_j, & \hfill\label{eq:confused-1}\\
\text{and} && \bm{x}_{N_G^+(j)} &= \bm{x}'_{N_G^+(j)}. & \label{eq:confused-2}
\end{align}
\end{definition}

Since message tuples corresponding to adjacent vertices cannot be mapped to the same codeword (otherwise, some receiver~$j$ cannot decode $X_j$ due to \eqref{eq:confused-1} and \eqref{eq:confused-2}), any proper colouring scheme gives an index code (where the colours map to distinct index codewords), and vice versa. Hence, the total number of distinct codewords required for encoding equals the number of colours in the colouring scheme. Consequently, 
\begin{equation}
r_{m^t}(G) = \frac{\log_2\chi(G_\text{confusion})}{\log_2|\mathcal{X}|},
\end{equation}
where $\chi(G)$ denotes the chromatic number of the undirected graph $G$. Note that the code here can be non-linear.

\begin{remark}
Using the method of confusion graph to determine $r_{m^t}(G)$ is intractable when the message alphabet size or the number of messages grows. Furthermore, this method alone cannot be used to determine $r(G)$.
\end{remark}

\subsubsection{5-cycle with binary messages}

If $\mathcal{G}$ is a 5-cycle (a member of $\mathcal{G}_\text{s}$) and the messages are binary, i.e., $|\mathcal{X}|=2$, its confusion graph $G_\text{confusion}$ contains 32 vertices and 240 edges. One can use a brute-force search to find that $\chi(G_\text{confusion}) = 8$. This gives $r_2(\mathcal{G})=3$~\cite{baryossefbirk11}. This corresponds to the light grey cell marked with an asterisk in Table~\ref{table:1}. For this case, it turns out that scalar linear codes are optimal.

\subsubsection{Other members in \textnormal{$\mathcal{G}_\text{s}$}}

For other members in $\mathcal{G}_\text{s}$, we first consider $G_\text{A}$ and $G_\text{B}$ in Figure~\ref{fig:special-group}. For these two graphs, we find that $\chi(G_\text{confusion}) = 7$. This means $r_2(G) = 2.8074$. The optimal codelength can be achieved by non-linear codes that map $\{0,1\}^5 \mapsto \{0,1,\dotsc,6\}$, where we choose the output alphabet size to be $|\mathcal{Y}|^p = 7$.

For the rest of the members in $\mathcal{G}_\text{s}$, one can repeat this procedure to calculate $r_2(G)$.

\subsubsection{Restricting the output alphabet to be a binary vector}

Now, if we restrict the output alphabet to be a binary vector, we have the following:
\begin{theorem} \label{theorem:g-s-binary}
For any $G \in \mathcal{G}_\text{s}$,
\begin{equation}
r_{m^t}(G) \leq 3,
\end{equation}
for any integers $m \geq 2$ and $t \geq 1$.
Furthermore, if $m=2$, $t=1$, and $|\mathcal{Y}|=2$, then
\begin{equation}
r_2(G) = 3,
\end{equation}
and the optimal index codelength is achievable using binary scalar linear codes.
\end{theorem}

\begin{IEEEproof}
\textit{(Achievability):} 
From Theorem~\ref{thm:5-others}, for any $G \in \mathcal{G}_\text{s}$, $\mais(G) = 2$.  We can always remove some arc(s) (dotted or solid) from $G$ to obtain a subgraph $G^-$ where $\mais(G^-)=3$ and $|V(G)^-|=5$. With this, we have
\begin{equation}
r_{m^t}(G) \leq r_{m^t}(G^-) = 3,
\end{equation}
for any $m \geq 2$, $t \geq 1$. Here, the inequality is due to Lemma~\ref{lem:remove-add-arcs} in Appendix~\ref{sec:lemma-proof}, and the equality follows from Theorem~\ref{thm:mais}. So, a scalar linear code of length~3 exists for $G$ for any $m$ and $t$.

\textit{(Lower bound):} 
We have manually found that $\chi(G_\text{confusion}') = 7$, where $G_\text{confusion}'$ is the confusion graph of any $G' \in \{G_\text{A}, G_\text{B}\}$ for $m=2$ and $t=1$. From the proof of Lemma~\ref{lem:remove-add-arcs} in Appendix~\ref{sec:lemma-proof}, for any $G \in \mathcal{G}_\text{s}$ with the corresponding confusion graph $G_\text{confusion}$, we have that
\begin{equation}
\chi(G_\text{confusion}) \geq \chi(G'_\text{confusion}) = 7. \label{eq:confusion-lower-bound}
\end{equation}

By definition,
\begin{equation}
r_{m^t}(G) = \frac{\log_2\chi(G_\text{confusion})}{\log_2|\mathcal{X}|} = \min_{\phi,\{\psi_i\}} \frac{p \log_2 |\mathcal{Y}|}{\log_2 2},
\end{equation}
where $p$ is the length of the codewords.

If we restrict the codeword to be binary vectors, i.e., $|\mathcal{Y}|=2$, we have
\begin{equation}
p \geq \log_2\chi(G_\text{confusion}) \geq \log_2 7 = 2.8074.
\end{equation}
Since $p$ must be an integer, we have $p \geq 3$. We complete the proof by noting the existence of length-3 scalar linear codes.
\end{IEEEproof}

\section{Conclusion}

In this paper, we have studied unicast index coding, a special class of index coding where each receiver requests only one message, and each message is requested by only one receiver. To find the optimal index codelength and optimal index codes, we have used a graphical approach of representing each index-coding instance by a directed graph. We first derived the optimal index codelength for all graphs whose order is at most two more than that of its maximum acyclic induced subgraph. We then used this result, combined with a combinatoric approach, to derive the optimal index codelength for all graphs with five or fewer vertices. We also showed that linear codes are optimal for all graphs in these two classes. While existing results give the optimal index codelength for all graphs with five or fewer vertices when the message alphabet size tends to infinity, in this work, we find the optimal codelength when the message alphabet size is finite.

\appendices

\section{Proof of Lemma~\ref{lemma:structure}: A Special Configuration} \label{appendix:structure}

Recall that $G$ must satisfy these two conditions:
\begin{itemize}
\item (\textsf{C1}) $\mais(G) = |V(G)| - 2$.
\item (\textsf{C2}) There are no two disjoint cycles in $G$.
\end{itemize}

From (\textsf{C1}) and (\textsf{C2}), $G$ must contain at least two cycles (which are not disjoint). 
Let $G_\text{sub}$ be the subgraph of $G$ induced by all vertices belonging to at least one cycle. It follows that $G_\text{sub}$ satisfies both conditions~(\textsf{C1}) and (\textsf{C2}).
We will proceed to show that $G_\text{sub}$ must contain the configuration in Figure~\ref{fig:three-cycles}. 
We will build the configuration from a cycle, say $C_1$, in $G_\text{sub}$. We call it the \emph{centre cycle}. We re-label the vertices in $G_\text{sub}$ such that the vertices in $C_1$ are in an ascending order in the direction of the arcs, i.e., $1 \rightarrow 2 \rightarrow \dotsm \rightarrow (|V(C_1)|-1)  \rightarrow |V(C_1)| \rightarrow 1$, where the choice of vertex~1 is arbitrary. 

For any path $P$ that originates from vertex $b$ and terminates at vertex $c$, i.e., $b \rightarrow \dotsm \rightarrow c$, we refer to all $\{z: z \in V(P) \setminus \{b, c\}\}$ as the \emph{internal vertices} of $P$. Here, we allow $b = c$; in such a case, $P$ is a cycle. We define the \textit{intersection} of two paths as the set of vertices common to them.

 We first show the following:
\begin{proposition} \label{proposition:outside-cycle}
Consider the subgraph $G_\text{sub}$ and the cycle $C_1$ in the subgraph. Every arc not in $C_1$ belongs to some \emph{outer path}, defined as a path that originates from a vertex in $C_1$ and terminates at a vertex (which can be the same vertex) in $C_1$, but with all arcs and all internal vertices (if exists) not in $C_1$.
\end{proposition}

\begin{IEEEproof}
Since $G_\text{sub}$ contains only cycles, any arc, say $(i \rightarrow j)$, not in $C_1$ must belong to another cycle, say $C_2$. Furthermore, from condition~(\textsf{C2}), $C_2$ must intersect $C_1$. Hence, $(i \rightarrow j)$ must belong to an outer path that originates from $C_1$ and terminates at $C_1$. 
\end{IEEEproof}

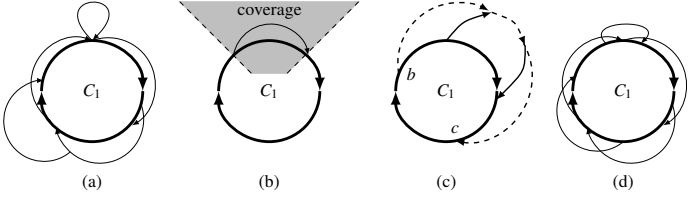
\begin{figure}[t]
\hspace*{-3mm}
\resizebox{9.3cm}{!}{%
\begin{tikzpicture}
\def\R{10mm}
\def\D{-1.8*\R}

\begin{scope}[xshift=0,yshift=0]
\node (0) at (0,0) {$C_1$};

\draw[ultra thick, ->, >=latex] (0:\R) arc (0:-180:\R);
\draw[ultra thick, ->, >=latex] (180:\R) arc (180:0:\R);

\draw[->,>=latex] (90:\R) .. controls +(-\R,\R)  and +(\R,\R) .. (90:\R);

\draw[->,>=latex] (90:\R) arc (110:-60:0.93*\R);
\draw[->,>=latex] (-15:\R) arc  (20:-170:0.87*\R) coordinate[pos=0.8] (C);

\draw[->,>=latex] (C) arc (310:90:0.8*\R);

\draw[<-,>=latex] (90:\R) arc (70:235:0.9*\R);

\node (label1) at (0,\D) {(a)};
\end{scope}

\begin{scope}[xshift=3.5*\R,yshift=0]
\node (0) at (0,0) {$C_1$};

\draw[fill, color=lightgray] (135:0.5*\R) -- (135:2.5*\R) -- (45:2.5*\R) -- (45:0.5*\R) -- cycle;
\draw[dashed] (135:0.5*\R) -- (135:2.5*\R);
\draw[dashed] (45:0.5*\R) -- (45:2.5*\R);

\node (label) at (90:1.55*\R) {coverage};

\draw[ultra thick, ->, >=latex] (0:\R) arc (0:-180:\R);
\draw[ultra thick, ->, >=latex] (180:\R) arc (180:0:\R);

\draw[->,>=latex] (135:\R) arc (170:10:0.75*\R);

\node (label2) at (0,\D) {(b)};

\end{scope}

\begin{scope}[xshift=7*\R,yshift=0]
\node (0) at (0,0) {$C_1$};

\draw[ultra thick, ->, >=latex] (0:\R) arc (0:-180:\R);
\draw[ultra thick, ->, >=latex] (180:\R) arc (180:0:\R);

\coordinate (a) at (60:1.8*\R);
\coordinate (b) at (30:1.8*\R);
\node[inner sep=0,label=right:$b$] (B) at (160:\R) {};
\node[inner sep=0,label=above:$c$] (C) at (-80:\R) {};

\draw[->,>=latex,thick] (90:\R) .. controls +(50:0.5*\R) .. (a);
\draw[->,>=latex,dashed,thick] (a) .. controls +(-20:0.5*\R) .. (b);
\draw[->,>=latex,thick] (b) .. controls +(-90:0.5*\R) .. (-10:\R);

\draw[->,>=latex,dashed, thick] (B) arc (190:60:1.2*\R);
\draw[<-,>=latex,dashed, thick] (C) arc (-100:30:1.3*\R);

\node (label3) at (0,\D) {(c)};

\end{scope}

\begin{scope}[xshift=10.5*\R,yshift=0]
\node (0) at (0,0) {$C_1$};

\draw[ultra thick, ->, >=latex] (0:\R) arc (0:-180:\R);
\draw[ultra thick, ->, >=latex] (180:\R) arc (180:0:\R);

\draw[->,>=latex] (110:\R) .. controls +(-0.5*\R,0.5*\R)  and +(0.5*\R,0.5*\R) .. (70:\R);

\draw[->,>=latex] (90:\R) arc (110:-60:0.93*\R);
\draw[->,>=latex] (-15:\R) arc (20:-170:0.87*\R);

\draw[->,>=latex] (-90:\R) arc (310:120:0.8*\R);

\draw[<-,>=latex] (90:\R) arc (70:235:0.9*\R);
\node (label4) at (0,\D) {(d)};
\end{scope}

\end{tikzpicture}
}%
\caption{We can always draw $G_\text{sub}$ as in subfigure~(a), i.e., a centre cycle $C_1$ and outer paths from $C_1$ and back to $C_1$. Subfigure~(b) shows the coverage of an outer path, i.e., vertices in $C_1$ in the grey area \emph{excluding} the two end points. Subfigure~(c) shows that when multiple outer paths originate from one vertex, we consider only the outer path with the largest coverage, i.e., the dashed path from $b$ to $c$. The outer paths in subfigure~(d) provide full coverage.}
\label{fig:redraw}
\end{figure}

It follows from Proposition~\ref{proposition:outside-cycle} that $G_\text{sub}$ consists of only a cycle $C_1$ and outer paths (from $C_1$ and back to $C_1$).
 Figure~\ref{fig:redraw}(a) shows an example of $G_\text{sub}$ where $C_1$ is marked with thick arrows and all outer paths thin arrows. Also, the outer paths cannot form any cycle outside $C_1$. Otherwise, we have two vertex-disjoint cycles, and this violates condition~(\textsf{C2}). 

We now prove a key proposition for proving Lemma~\ref{lemma:structure}.
\begin{proposition} \label{proposition:outer-path}
Remove vertex~1 in $C_1$. There exists another cycle in $G_\text{sub}$ if and only if there is an outer path from some $b \in V(C_1) \setminus \{1\}$ to some $c \in V(C_1) \setminus \{1\}$, where $b \geq c$.
\end{proposition}

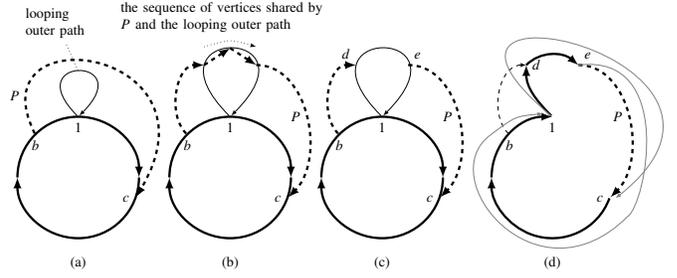
\begin{figure}[t]
\hspace*{-5mm}
\resizebox{11cm}{!}{%
\begin{tikzpicture}
\linespread{1}
\def\R{15mm}
\begin{scope}[xshift=0,yshift=0]
\draw[->,>=latex, ultra thick] (-\R,0) arc (180:0:\R); 
\draw[->,>=latex, ultra thick] (\R,0) arc (360:180:\R); 
\node [inner sep = 0,label=below: {$1$}] (1) at (0,\R) {};
\draw[->,>=latex] (1) .. controls +(-\R,\R)  and +(\R,\R) .. coordinate[pos=0.5] (A) (1);
\node [inner sep = 0,label=below: {$b$}] (b) at (135:\R) {};
\node [inner sep = 0,label=left: {$c$}] (c) at (-20:\R) {};
\draw[ultra thick, ->,>=latex,dashed] (b) .. controls +(-\R,\R*2) and +(\R*1.5,\R*2.5) .. node[very near start,left] {$P$} (c);
\node[inner sep=0,label=above:{\parbox[t]{2cm}{looping\\ outer path}}] (l) at  (-0.2*\R,2.2*\R) {};
\draw [dotted] (l) -- (A);
\node (label1) at (0,-1.4*\R) {(a)};
\end{scope}

\begin{scope}[xshift={2.5*\R},yshift=0]
\draw[ultra thick,->,>=latex] (-\R,0) arc (180:0:\R); 
\draw[ultra thick,->,>=latex] (\R,0) arc (360:180:\R); 
\node [inner sep = 0,label=below: {$1$}] (1) at (0,\R) {};
\draw[->,>=latex] (1) .. controls +(-1.5*\R,1.5*\R)  and +(1.5*\R,1.5*\R) .. coordinate[pos=0.25] (Ca) coordinate[pos=0.5] (Da) coordinate[pos=0.75] (Ea) (1);
\node [inner sep = 0,label=below: {$b$}] (b) at (135:\R) {};
\node [inner sep = 0,label=left: {$c$}] (c) at (-20:\R) {};

\draw[ultra thick,->,>=latex,dashed] (b) to [out=135,in=180] (Ca);
\draw[ultra thick,->,>=latex,dashed] (Ca) -- (Da);
\draw[ultra thick,->,>=latex,dashed] (Da) -- (Ea);
\draw[ultra thick,->,>=latex,dashed] (Ea) to [out=0,in=45] node[midway,left] {$P$} (c);

\coordinate[yshift=0.3*\R] (Caup) at (Ca);
\coordinate[yshift=0.2*\R] (Daup) at (Da);
\coordinate[yshift=0.3*\R] (Eaup) at (Ea);

\node[yshift=-0.1*\R,xshift=0.2*\R,label=above:{\parbox[t]{6cm}{the sequence of vertices shared by\\ $P$ and the looping outer path}}] () at (Daup) {};

\draw[->,>=latex,dotted] (Caup) .. controls ([yshift=-0.05*\R]Daup) .. (Eaup);

\node (label2) at (0,-1.4*\R) {(b)};
\end{scope}

\begin{scope}[xshift={5*\R},yshift=0]
\draw[ultra thick,->,>=latex] (-\R,0) arc (180:0:\R); 
\draw[ultra thick,->,>=latex] (\R,0) arc (360:180:\R); 
\node [inner sep = 0,label=below: {$1$}] (1) at (0,\R) {};
\draw[->,>=latex] (1) .. controls +(-1.5*\R,1.5*\R)  and +(1.5*\R,1.5*\R) .. coordinate[pos=0.25] (Ca) coordinate[pos=0.75] (Ea) (1);
\node [inner sep = 0,label=below: {$b$}] (b) at (135:\R) {};
\node [inner sep = 0,label=left: {$c$}] (c) at (-20:\R) {};

\draw[ultra thick,->,>=latex,dashed] (b) to [out=135,in=180] (Ca);
\draw[ultra thick,->,>=latex,dashed] (Ea) to  [out=0,in=45] node[midway,left] {$P$} (c);
\node[inner sep=0, label=north west:$d$] (Cb) at (Ca) {};
\node[inner sep = 0,label=north east:$e$] (Eb) at (Ea) {};
\node (label3) at (0,-1.4*\R) {(c)};

\end{scope}

\begin{scope}[xshift={7.8*\R},yshift=0]
\draw[ultra thick,->,>=latex] (-\R,0) arc (180:90:\R); 
\draw[ultra thick,->,>=latex] (-20:\R) arc (-20:-180:\R); 
\node [inner sep = 0,label=below: {$1$}] (1) at (0,\R) {};
\path[ultra thick,->,>=latex] (1) .. controls +(-1.5*\R,1.5*\R)  and +(1.5*\R,1.5*\R) .. coordinate[pos=0.25] (Ca) coordinate[pos=0.75] (Ea) (1);
\node [inner sep = 0,label=below: {$b$}] (b) at (135:\R) {};
\node [inner sep = 0,label=left: {$c$}] (c) at (-20:\R) {};

\coordinate (Bdup) at (b);
\coordinate (Cdup) at (c);
\coordinate (1dup) at (1);

\draw[->,>=latex,dashed ] (b) to [out=135,in=180] (Ca);
\draw[ultra thick,->,>=latex,dashed] (Ea) to  [out=0,in=45] node[midway,left] {$P$} (c);
\node[inner sep=0, label=east:$d$] (Cb) at (Ca) {};
\node[inner sep = 0,label=north east:$e$] (Eb) at (Ea) {};
\draw[ultra thick, ->, >=latex] (1) to [out=135,in=270] (Ca);
\draw[ultra thick, ->, >=latex] (Ca) to [out=45,in=140] (Ea);


\draw[->, >=latex, color=gray] (1) .. controls (-3*\R,3.5*\R) and (4*\R,2*\R) .. ([xshift=0.1*\R]Cdup);

\draw[->, >=latex, color=gray] (Ea) .. controls +(1.6*\R,0.2*\R) and +(0.3*\R,0.2*\R)  ..  ([yshift=-0.3*\R,xshift=0.3*\R]Cdup) .. controls +(-1.5*\R,-1.5*\R) and (-2*\R,0) .. ([xshift=-0.3*\R]Bdup) .. controls +(0.4*\R,0.3*\R) .. ([yshift=0.05*\R,xshift=-0.15*\R]1dup);

\node (label4) at (0,-1.4*\R) {(d)};

\end{scope}

\end{tikzpicture}
}
\vspace*{-5ex}
\caption{Case 1 where there exists a looping outer path (drawn with thin solid lines) that starts and ends at vertex~1. The centre cycle $C_1$ is drawn with thick solid lines, and the second outer path (denoted as $P$) from $b$ to $c$, dashed lines. To get another cycle after removing vertex~1, we must have that $1 < c \leq b \leq |V(C_1)|$, as shown in subfigure~(a). However, there are two vertex-disjoint cycles in subfigure~(a). So, $P$ must intersect the looping outer path, as shown in subfigure (b). Taking the segment of $P$ from $C_1$ to the looping outer path, and that from the looping outer path back to $C_1$, we have subfigure (c). We can re-draw the path from $1$ to $c$ and that from $e$ to $1$ in subfigure (c) to get subfigure (d), where we have drawn the new centre cycle with thick (both solid and dashed) lines.}
\label{fig:looping}
\end{figure}

\begin{IEEEproof}
{[}The converse:] We remove vertex~1. If there is another cycle, then there is a vertex (not vertex~1) in $C_1$ that has a path back to itself (this is because any cycle must share some vertex with $C_1$). This cannot happen if every outer path terminates at a higher-indexed vertex (we can ignore all outer paths that originate or terminate at vertex~1 as the vertex has been removed). So, there must exist an outer path with $b \geq c$.

[The forward part:] Clearly, if $b = c$, we have another cycle formed by the outer path. Otherwise, i.e., $ b > c$, the outer path and the path along $C_1$ from $c$ to $b$ form a cycle. See Figure~\ref{fig:looping}(a) for an example.
\end{IEEEproof}

Next, we define a \emph{looping} outer path as an outer path that originates and terminates at the same vertex in $C_1$. The graph $G_\text{sub}$ can be categorised as follows:
\begin{itemize}
\item there exists at least one looping outer path (Case 1), or
\item there is no looping outer path (which we will further divide into Cases 2 and 3).
\end{itemize}

We will show that in any case, we have Figure~\ref{fig:three-cycles}.

\subsection{Case 1: There exists a looping outer path}
Suppose that there exists a looping outer path from and to vertex~$1 \in V(C_1)$. This incurs no loss of generality as the choice of vertex~1 is arbitrary. Removing vertex~1 disconnects both cycle $C_1$ and the cycle formed by the looping outer path. Recall that we need to remove two vertices to disconnect all cycles in $G_\text{sub}$. So, there must exist another cycle in $G_\text{sub}$. 

From Proposition~\ref{proposition:outer-path}, there exists another outer path $P$ from $b \in V(C_1) \setminus \{1\}$ to $c \in V(C_1) \setminus \{1\}$, where $b \geq c$. The outer path $P$ must intersect the looping outer path; otherwise there exist two cycles as shown in Figure~\ref{fig:looping}(a).

Re-label the internal vertices of the looping outer path in an ascending order, as follows: $1 \rightarrow (|V(C_1)|+1) \rightarrow (|V(C_1)|+2) \rightarrow \dotsm \rightarrow (|V(C_1)|+L) \rightarrow 1$, where $L$ is the number of internal vertices in the looping outer path. It follows that the intersection of $P$ and the looping outer path must be in an ascending order in the direction of $P$ (see Figure~\ref{fig:looping}(b)); otherwise, a cycle forms outside $C_1$.

See Figure~\ref{fig:looping}(c). Consider only the following segments of $P$: \begin{enumerate}
  \item from $b$ to the vertex where $P$ first intersect the looping outer path, denoted by $d$ (which is the smallest vertex index in the intersection set); and
  \item the vertex where $P$ leaves the looping outer path, denoted by $e$ (which is the largest vertex index in the intersection set), to $c$.
  \end{enumerate}
  It follows that $d \leq e$. By construction, all paths in Figure~\ref{fig:looping}(c) intersect only at vertices $1$, $b$, $c$, $d$, and $e$. 
Finally, re-draw Figure~\ref{fig:looping}(c) to get Figure~\ref{fig:looping}(d), which is isomorphic to  Figure~\ref{fig:three-cycles} (where the thick lines in Figure~\ref{fig:looping}(d) correspond to paths $I$, $H$, $U$, $C$, $W$, and $D$ in Figure~\ref{fig:three-cycles}).

Note that vertices~1, $b$, and $d$ must be unique. 
We have shown that if there is a looping outer path, then we have the configuration in Figure~\ref{fig:three-cycles}, where path $I$ has zero arc, paths $W$ and $U$ possibly have zero arc (if $b=c$ and/or $d=e$), and all other paths must contain at least one arc.

\subsection{No looping outer path}

For a non-looping outer path from vertex $b \in V(C_1)$ to $c \in V(C_1) \setminus \{b\}$, we say that the vertices \underline{in $C_1$} from $b$ to $c$ (in the direction of the arcs in $C_1$) but excluding $b$ and $c$ is \emph{covered} by this outer path. See Figure~\ref{fig:redraw}(b) for an example. 

For the analyses in this paper, we exclude all outer paths with strictly smaller coverage, or multiple outer paths with equal coverage. Referring to Figure~\ref{fig:redraw}(c), consider an outer path that originates from $b$. Suppose that it has multiple paths back to $C_1$. We consider only the path (back to $C_1$) that has the \emph{largest coverage}. Similarly, for any path that terminates at $c$, we consider only the path (leaving $C_1$) that has the largest coverage. By doing this, no two outer paths have the same originating vertex or terminating vertex. 

We now show the following property:
\begin{proposition} \label{proposition:full-coverage}
If there is no looping outer paths in $G_\text{sub}$, then all largest-covering outer paths must, together, provide full coverage for the cycle $C_1$. In other words, every vertex in $C_1$ must be covered by at least one outer path.
\end{proposition}

\begin{IEEEproof}
Consider any vertex $a \in V(C_1)$. Re-label $a$ as vertex~1, and other vertices $V(C_1)$ in an ascending order in the arc direction. Remove vertex~1 from $G_\text{sub}$. There must exist another cycle.  It follows from Proposition~\ref{proposition:outer-path} that an outer path $P$ from $b$ to $c$ must exist, where $1 < c < b \leq |V(C_1)|$  ($c \neq b$ since there is no looping path), meaning that this outer path must cover vertex~1. We can safely ignore other outer paths that provide smaller or equal coverage, because if $P$ does not cover vertex~1, then none of the ignored outer paths does. Since the choice of $a$ is arbitrary, we have Proposition~\ref{proposition:full-coverage}.
\end{IEEEproof}

For example, the outer paths in Figure~\ref{fig:redraw}(d) provide full coverage for $C_1$, but the outer paths in Figures~\ref{fig:redraw}(a)--(c) do not. Removing any uncovered vertex from $C_1$ makes $G_\text{sub}$ acyclic.

From here, we consider $G_\text{sub}$ that consists of cycle $C_1$ and all outer paths that provide the largest coverage (i.e., we remove all other arcs and vertices).  We are ready to proceed with Cases 2 and 3, defined as follows:
\begin{itemize}
\item (Case 2) There is no looping outer path, and no two outer paths have any common internal vertex.
\item (Case 3) There is no looping outer path, and there exist two outer paths sharing the same internal vertex.
\end{itemize}

\subsection{Case 2: No looping outer path, and all outer paths do not share internal vertices} 
We will show that we can always find three outer paths that provide full coverage.

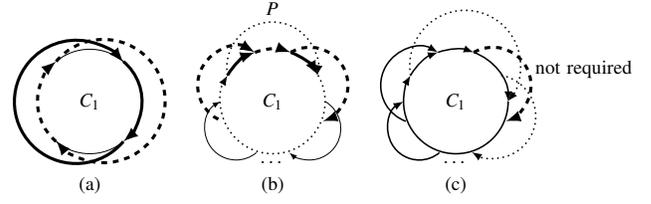
\begin{figure}[t]
\centering
\resizebox{8.5cm}{!}{%
\begin{tikzpicture}
\def\R{6ex}
\def\D{-1.6*\R}

\begin{scope}[xshift=0,yshift=0]
\node (0) at (0,0) {$C_1$};

\draw[ultra thick, ->, >=latex] (50:\R) arc (50:-50:\R);

\draw (50:\R) arc (50:130:\R);
\draw (-50:\R) arc (-50:-130:\R);

\draw[ultra thick, ->, >=latex] (-50:\R) arc (-40:-320:1.18*\R);

\draw[dashed, ultra thick, ->, >=latex] (230:\R) arc (230:130:\R);

\draw[dashed, ultra thick, ->, >=latex] (130:\R) arc (140:-140:1.18*\R);

\node (label1) at (0,\D) {(a)};
\end{scope}

\begin{scope}[xshift=3.5*\R,yshift=0]

\node (0) at (0,0) {$C_1$};

\draw[dotted, thick, ->, >=latex] (30:\R) arc (30:-210:\R);

\draw[dotted, thick, ->, >=latex] (150:\R) arc (190:-10:0.87*\R) node[midway,above] {$P$};

\draw[dashed, ultra thick, ->, >=latex] (110:\R) arc (110:70:\R);

\draw[dashed, ultra thick, ->, >=latex] (70:\R) arc (120:-70:0.73*\R);
\draw[dashed, ultra thick, <-, >=latex] (110:\R) arc (60:250:0.73*\R);

\draw[ultra thick, ->, >=latex] (70:\R) arc (70:30:\R);
\draw[ultra thick, ->, >=latex] (150:\R) arc (150:110:\R);

\draw[->, >=latex] (0:\R) arc (60:-130:0.6*\R);
\draw[<-, >=latex] (180:\R) arc (120:310:0.6*\R);

\node (dots) at (0,-1.2*\R) {$\dotsm$};

\node (label2) at (0,\D) {(b)};

\end{scope}

\begin{scope}[xshift=7*\R,yshift=0]

\node (0) at (0,0) {$C_1$};

\draw[thick, ->, >=latex] (0:\R) arc (0:-210:\R);

\draw[dotted, thick, ->, >=latex] (150:\R) arc (190:-40:1.06*\R);

\draw[thick, ->, >=latex] (110:\R) arc (110:70:\R);

\draw[dashed, ultra thick, ->, >=latex] (70:\R) arc (120:-70:0.73*\R) node[midway,right] {not required};
\draw[thick, <-, >=latex] (110:\R) arc (60:250:0.73*\R);

\draw[thick, ->, >=latex] (70:\R) arc (70:0:\R);
\draw[thick, ->, >=latex] (150:\R) arc (150:110:\R);

\draw[dotted, thick, ->, >=latex] (30:\R) arc (80:-122:0.8*\R);
\draw[thick, <-, >=latex] (180:\R) arc (120:310:0.6*\R);

\node (dots) at (0,-1.2*\R) {$\dotsm$};

\node (label3) at (0,\D) {(c)};

\end{scope}
\end{tikzpicture}
}
\caption{(a) If two outer paths provide full coverage, we can always form two disjoint cycles, one formed by the thick solid path, and another one by the dashed path. (b) $G_\text{sub}$ with $K$ outer paths, where $K \geq 4$, providing full coverage can be converted to $K-2$ outer paths providing full coverage. (c) If two non-adjacent outer paths give overlapping coverage (e.g., the two dotted paths), then the paths in between are redundant (the dashed path), i.e., $K-1$ outer paths are sufficient to give full coverage, instead of $K$.}
\label{fig:2-coverage}
\end{figure}

First, note that one outer path cannot provide full coverage.
Suppose that we can find two outer paths providing full coverage. We illustrate in Figure~\ref{fig:2-coverage}(a) that we can always form two vertex-disjoint cycles. So, this scenario cannot happen.

Next, suppose that we can find three outer paths providing full coverage, we have exactly Figure~\ref{fig:three-cycles}. As there is no looping outer path, the nine paths in Figure~\ref{fig:three-cycles} each have one or more arcs.


Finally, we show that if we can find $K \geq 4$ outer paths providing full coverage, we can always modify the cycles such that $(K-2)$ outer paths provide full coverage. We illustrate this in Figure~\ref{fig:2-coverage}(b). We do the following:
\begin{enumerate}
\item Combine the two dotted paths to be the new $C_1$.
\item Combine the  two outer paths adjacent to $P$ (dashed paths) and the dashed path in $C_1$ that connects the two dashed outer paths (which can be of length 0) into a new outer path.
\item Remove all arcs and internal vertices in the the thick solid paths in $C_1$. 
\end{enumerate}
Note that by doing this, the new graph still retains the structure of a cycle with outer paths covering it. The new graph has $K-2$ outer paths providing full coverage. This reduction is always possible. This can be seen from Figure~\ref{fig:2-coverage}(b), where the outer path to the left of $P$ must terminate within the coverage of $P$, and that to the right of $P$ must start within the coverage of $P$. Otherwise, they cannot provide full coverage to $C_1$. Also, within the coverage of $P$, the coverage areas of the two adjacent paths do not overlap. Otherwise, $P$ is not required, as illustrated in Figure~\ref{fig:2-coverage}(c).

By repeating this step, starting from $K$ outer paths, for any $K \geq 4$, we can find a graph with $K=2$ or $K=3$ outer paths. As $K=2$ is not possible, we will always get a graph with $K=3$ outer paths providing full coverage, which is in the form of Figure~\ref{fig:three-cycles}.

\subsection{Case 3: No looping outer path and two outer paths share some internal vertices} \label{sec:figure-end}
Let the two outer paths that share some common internal vertex be $P$ and $Q$, and one of the shared internal vertices be $z$.  Further, let the originating and terminating vertices of $P$ be $p_1$ and $p_\text{last}$ respectively, and those of $Q$ be $q_1$ and $q_\text{last}$. Here, $p_1 \neq p_\text{last}$ and $q_1 \neq q_\text{last}$ as there is no looping outer path, and $p_1 \neq q_1$ and $p_\text{last} \neq q_\text{last}$ as no two outer paths have the same originating or terminating vertex.

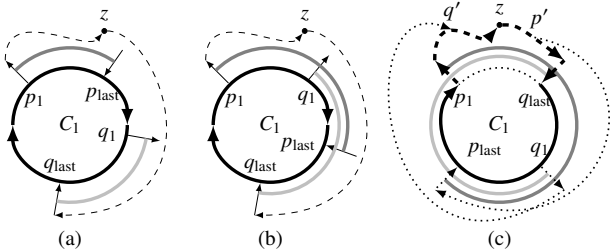
\begin{figure}[t]
\centering
\resizebox{9cm}{!}{%
\begin{tikzpicture}
\def\R{6ex}
\def\D{-2*\R}
\def\S{1.6*\R}
\def\U{1.2*\R}
\def\T{1.35*\R}

\begin{scope}[xshift=0,yshift=0]
\node (0) at (0,0) {$C_1$};

\draw[ultra thick, ->, >=latex] (0:\R) arc (0:-180:\R);
\draw[ultra thick, ->, >=latex] (180:\R) arc (180:0:\R);

\node[inner sep=0,label=south:$\,\,\,p_1$] (p1) at (135:\R) {};
\node[inner sep=0,label=south:$p_\text{last}$] (plast) at (55:\R) {};
\node[inner sep=0,label=west:$q_1$] (q1) at (-10:\R) {};
\node[inner sep=0,label=north:$q_\text{last}$] (qlast) at (-100:\R) {};

\draw[->,>=latex] (p1) -- (135:\S);
\draw[<-,>=latex] (plast) -- (55:\S);
\draw[->,>=latex] (q1) -- (-10:\S);
\draw[<-,>=latex] (qlast) -- (-100:\S);

\draw[color=gray, ultra thick] (135:\T) arc (135:55:\T);
\draw[color=lightgray, ultra thick] (-10:\T) arc (-10:-100:\T);

\node[inner sep = 0.03*\R,draw, fill,circle, label=above:$z$] (z) at  (70:1.75*\R) {};

\draw[dashed,->,>=latex] (135:\S) .. controls +(90:\R) and +(-130:0.5*\R) .. (z);
\draw[dashed,->,>=latex] (z) .. controls +(0:\R) and +(0:3*\R) .. (-100:\S);

\node (label1) at (0,\D) {(a)};
\end{scope}

\begin{scope}[xshift=3.5*\R,yshift=0]
\node (0) at (0,0) {$C_1$};

\draw[ultra thick, ->, >=latex] (0:\R) arc (0:-180:\R);
\draw[ultra thick, ->, >=latex] (180:\R) arc (180:0:\R);

\node[inner sep=0,label=south:$\,\,\,p_1$] (p1) at (135:\R) {};
\node[inner sep=0,label=west:$p_\text{last}$] (plast) at (-20:\R) {};
\node[inner sep=0,label=south:$q_1$] (q1) at (50:\R) {};
\node[inner sep=0,label=north:$q_\text{last}$] (qlast) at (-100:\R) {};

\draw[->,>=latex] (p1) -- (135:\S);
\draw[<-,>=latex] (plast) -- (-20:\S);
\draw[->,>=latex] (q1) -- (50:\S);
\draw[<-,>=latex] (qlast) -- (-100:\S);

\draw[color=gray, ultra thick] (135:\T) arc (135:-20:\T);
\draw[color=lightgray, ultra thick] (50:\U) arc (50:-100:\U);

\node[inner sep = 0.03*\R,draw, fill,circle, label=above:$z$] (z) at  (70:1.75*\R) {};

\draw[dashed,->,>=latex] (135:\S) .. controls +(90:\R) and +(-130:0.5*\R) .. (z);
\draw[dashed,->,>=latex] (z) .. controls +(0:\R) and +(0:3*\R) .. (-100:\S);

\node (label2) at (0,\D) {(b)};
\end{scope}

\begin{scope}[xshift=7.5*\R,yshift=0]
\node (0) at (0,0) {$C_1$};

\draw[ultra thick, ->, >=latex] (45:\R) arc (45:-225:\R);
\draw[dotted, thick] (45:\R) arc (45:135:\R);

\node[inner sep=0,label=south:$\,\,\,p_1$] (p1) at (135:\R) {};
\node[inner sep=0,label=north east:$p_\text{last}$] (plast) at (-135:\R) {};
\node[inner sep=0,label=north:$q_1$] (q1) at (315:\R) {};
\node[inner sep=0,label=south:$q_\text{last}\,\,\,$] (qlast) at (45:\R) {};

\draw[color=gray, ultra thick] (135:\T) arc (135:-135:\T);
\draw[color=lightgray, ultra thick] (315:\U) arc (315:45:\U);

\node[inner sep = 0.03*\R,draw, fill,circle, label=above:$z$] (z) at  (90:1.75*\R) {};

\draw[dashed,->,>=latex, ultra thick] (135:\S) .. controls +(90:\R) and +(-130:0.5*\R) .. (z) coordinate[pos=0.4] (A2);
\draw[dashed,->,>=latex, ultra thick] (z) .. controls +(0:0.3*\R) .. (45:\S) coordinate[pos=0.7] (B2);

\draw[->,>=latex, dashed, ultra thick] (p1) -- (135:\S);
\draw[<-,>=latex, dotted, thick] (plast) -- (-135:\S);
\draw[->,>=latex, dotted, thick] (q1) -- (315:\S);
\draw[<-,>=latex, dashed, ultra thick] (qlast) -- (45:\S);

\draw[dotted,thick,->,>=latex] (315:\S) .. controls +(-145:4*\R) and +(160:2*\R) .. (A2);

\draw[dotted,thick,<-,>=latex] (-135:\S) .. controls +(-25:3.3*\R) and +(-10:2.5*\R) .. (B2);

\node[inner sep=0,label=above:$q'$] (qprime) at (A2) {};
\node[inner sep=0,label=above:$\,\,\,\,p'$] (pprime) at (B2) {};

\node (label3) at (0,\D) {(c)};
\end{scope}

\end{tikzpicture}
}%
\vspace*{-5ex}
\caption{The overlapping of the coverage of two outer paths, where the dark grey lines represent the coverage of outer path $P$ ($p_1 \rightarrow \dotsm \rightarrow p_\text{last}$), and the light grey lines that of outer path $Q$ ($q_1 \rightarrow \dotsm \rightarrow q_\text{last}$)}
\label{fig:overlapping-coverage}
\end{figure}

Now, the coverage of $P$ and $Q$ can be either (a) non-overlapping, (b) overlapping once, or (c) overlapping twice, as shown in Figure~\ref{fig:overlapping-coverage}. The dark grey line shows the coverage of $P$, and the light grey line that of $Q$. By definition, there is a subpath from $p_1$ to $z$ along $P$ and another subpath from $z$ to $p_\text{last}$ along $P$. The two subpaths must be vertex-disjoint, except $z$, as there is no cycle in $P$.  Similarly, we have two vertex-disjoint paths from $q_1$ to $z$, and from $z$ to $q_\text{last}$, both along $Q$. This means, there is an subpath from $p_1$ to $q_\text{last}$ through $z$, and another from $q_1$ to $p_\text{last}$ through $z$. So, $p_1 \neq q_\text{last}$, $q_1 \neq p_\text{last}$, as there is no looping outer path, and hence $p_1$, $p_\text{last}$, $q_1$, and $q_\text{last}$ are distinct. 

Suppose that we have Figure~\ref{fig:overlapping-coverage}(a). The largest-covering outer path from $p_1$ should terminate at $q_\text{last}$, and that from $q_1$ at $p_\text{last}$. The outer path from $p_1$ to $q_\text{last}$ and that from $q_1$ to $p_\text{last}$ should have been chosen. This means the largest-covering paths actually overlap twice, i.e., we should have Figure~\ref{fig:overlapping-coverage}(c).

Suppose that we have Figure~\ref{fig:overlapping-coverage}(b). The outer path from $p_1$ to $q_\text{last}$, through $z$, gives the largest coverage, and it would have been chosen. 

So, we can only have the configuration in Figure~\ref{fig:overlapping-coverage}(c), where the coverage of $P$ and $Q$ overlaps twice. The coverage from $p_1$ to $q_\text{last}$ is smaller than that from $p_1$ to $p_\text{last}$. So, the largest-covering outer path from $p_1$ was correctly identified. Similarly, the largest-covering outer path from $q_1$ terminates at $q_\text{last}$.


We will now show that we can always get Figure~\ref{fig:three-cycles} from Figure~\ref{fig:overlapping-coverage}(c). Recall that there is a subpath from $p_1$ to $z$ and another subpath from $z$ to $q_\text{last}$, and these two subpaths are vertex-disjoint, except $z$. Otherwise, we get a cycle disjoint from $C_1$. We denote the outer path from $p_1$ to $q_\text{last}$ (through $z$) by $Z$ (drawn with a thick dashed line). 

Next, recall that there is a subpath from $q_1$ to $z$, and another from $z$ to $p_\text{last}$. So, the subpath from $q_1$ to $z$ must intersect $Z$. Denote the vertex it first intersect $Z$ as $q'$. Similarly, the subpath from $z$ to $p_\text{last}$ must intersect $Z$ (at least at vertex~$z$). Let the last vertex in the intersection set be $p'$. With this construction, $Z$, the subpath from $q_1$ to $q'$, and the subpath from $p'$ to $p_\text{last}$ are vertex-disjoint, except at  $p'$ and $q'$.

We now re-draw Figure~\ref{fig:overlapping-coverage}(c) as follows: Let the path from $q_\text{last}$ to $p_1$ along $C_1$ (drawn with a thick solid line) plus path $Z$ (drawn with a thick dashed line) be the centre cycle, and let the subpaths (drawn with dotted paths) (i) from $p_1$ to $q_\text{last}$ along $C_1$, (ii) from $p'$ to $p_\text{last}$, and (iii) from $q_1$ to $q'$ be the three outer paths. Note that only $p'$ and $q'$ can co-locate. The resultant graph is isomorphic to Figure~\ref{fig:three-cycles}, with path $I$ possibly having zero arc (if $p' = q' = z$).

Combining the Cases 1--3, we have Lemma~\ref{lemma:structure}. $\hfill\blacksquare$

\section{Proof of Lemma~\ref{proposition:last-case}} \label{sec:lemma-proof}

We first note the following:

\begin{observation}
If $\mais(G)=2$, then
any induced subgraph of $G$ with three vertices must contain a cycle.\footnote{Recall that, unless stated otherwise, cycles refer to directed cycles.}
Otherwise, $\mais(G) \geq 3$ by considering the 3-vertex induced subgraph without a cycle.
\end{observation}

We define edges in directed graphs as follows:
\begin{definition}
Consider a directed graph $G$ with vertex set $V(G)$ and arc set $A(G)$. For a pair of vertices $i,j \in V(G)$, we say that there is an \textit{edge} between these two vertices if and only if  $(i \rightarrow j) \in A(G)$ and $(j \rightarrow i) \in A(G)$. A cycle formed by edges is called an \textit{undirected cycle}.
\end{definition} 

As the proof of the lemma is rather involved, we divide the set of all graphs to be considered in this lemma, i.e., all $G$ with $|V(G)|=5$, $G \notin \mathcal{G}_\text{s}$, and $\mais(G)=2$, into four categories according to the number of undirected cycles in $G$:
\begin{enumerate}
\item There is no undirected cycle.
\item There exists an undirected cycle of length 3.
\item There is no undirected cycle of length 3, but there exists an undirected cycle of length 4.
\item There is no undirected cycle of length 3 or 4, but there exists an undirected cycle of length 5.
\end{enumerate}

Note that, by definition, there cannot be any undirected cycle of length 2 or less.

\subsection{Two useful lemmas}

We say that $G^-$ is an \textit{arc-deleted} subgraph of $G$ if $V(G^-) = V(G)$ and $A(G') \subseteq A(G)$, i.e., removing zero or some arc(s) from $G$ but retaining all the vertices.

We first prove two lemmas to be used subsequently:
\begin{lemma} \label{lem:remove-add-arcs}
Let $G$ be an arc-deleted subgraph of $G^+$, and $G^-$ be an arc-deleted subgraph of $G$. Then,
\begin{equation}
r_{m^t}(G^+) \leq r_{m^t}(G) \leq r_{m^t}(G^-),
\end{equation}
and an index code for $G^-$ is an index code for $G$ and $G^+$.
\end{lemma}

\begin{IEEEproof}
Each receiver in $G^+$ has prior messages of at least what it has in $G$, and requests the same message (i.e., receiver~$i$ requests $X_i$). So, any index code for $G$ satisfies all decoding requirements for $G^+$ and hence is an index code for $G^+$. This proves $r_{m^t}(G^+) \leq r_{m^t}(G)$. By repeating the same argument, we have $r_{m^t}(G) \leq r_{m^t}(G^-)$.
\end{IEEEproof}

\begin{lemma} \label{lem:4-cycle}
If $|V(G)|=5$ and $\mais(G)=2$, then the induced subgraph of any four vertices must contain an edge. 
\end{lemma}

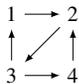
\begin{figure}[t]
\centering
\resizebox{1.25cm}{!}{%
\begin{tikzpicture}
\graph {
{[edge={>=latex}]
1 -> 2, 
3 -> 4,
4 -> 2 -> 3 -> 1
}
};
\end{tikzpicture}
}%

\caption{If there is no edge in the graph, then $\{1,3,4\}$ cannot contain a cycle. }
\label{fig:4-cycle}
\end{figure}

\begin{IEEEproof}
We will prove the lemma by contradiction. Suppose that there is an induced subgraph of four vertices without an edge.
Recall that any induced subgraph of three vertices must contain a cycle. Referring to Figure~\ref{fig:4-cycle}, there must be a directed cycle in $\{1,2,3\}$. Since there is no edge, there cannot be any 2-cycle. Without loss of generality, let the cycle be $1 \rightarrow 2 \rightarrow 3 \rightarrow 1$. Again, as there cannot be any edge, the cycle in $\{2,3,4\}$ must be $2 \rightarrow 3 \rightarrow 4 \rightarrow 2$. Now, for $\{1,3,4\}$ to contain a cycle, it must contain an an edge (contradiction). We would have obtained the same result had we started by choosing the cycle in $\{1,2,3\}$ to be $1 \rightarrow 3 \rightarrow 2 \rightarrow 1$.
\end{IEEEproof}

\subsection{Basic ideas}

We will prove Lemma~\ref{proposition:last-case} using the following ideas: For each category, we will show that any $G$ must contain some a arc-deleted subgraph, say $G_\text{sub}$. We then show that there exists a scalar linear index code of length 2 (over the ring $\mathcal{X}$) for $G_\text{sub}$, thereby establishing $r_{m^t}(G_\text{sub}) \leq 2$. Since $G_\text{sub} = G^-$, from Lemma~\ref{lem:remove-add-arcs}, we must have that $r_{m^t}(G) \leq 2$, where the 2-bit achievability uses the same linear code for $G_\text{sub}$. As $\mais(G)=2$ is a lower bound on $r_{m^t}(G)$, we establish $r_{m^t}(G)=2$. We will use a combinatoric approach.


\subsection{Category 1: No undirected cycle}

We start with the first category where there cannot be any undirected cycle in $G$. We have the following subcategories:

\subsubsection{There is one or no edge} 
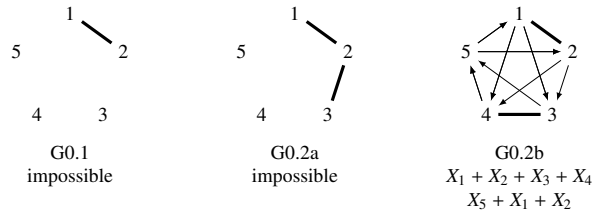
\begin{figure}[t]
\centering
\resizebox{80mm}{!}{%
\begin{tikzpicture}
\linespread{1}
\def\D{7ex}
\graph [clockwise, n=5] {
{[clique, edges=ultra thick] 1, 2},
3,
4,
5
};
\node[label={[align=center]below:G0.1\\impossible}] at (0,{-\D}) {};

\begin{scope}[xshift=40mm]
\graph [clockwise, n=5] {
{[path, edges=ultra thick] 1, 2, 3},
4,
5
};
\node[label={[align=center]below:G0.2a\\impossible}] at (0,{-\D}) {};
\end{scope}

\begin{scope}[xshift=80mm]
\graph [clockwise, n=5] {
{[path, edges={ultra thick}] 1, 2},
{[path, ->, edges={>=latex}] 2, 3},
{[path, edges=ultra thick] 3, 4},
{[path, ->, edges={>=latex}] 5 -> 1 -> 3 -> 5}, 
{[path, ->, edges={>=latex}]4 -> 5 -> 2 -> 4}, 
{[path, ->, edges={>=latex}]1-> 4}
};
\node[label={[align=center]below:G0.2b\\$X_1+X_2+X_3+X_4$\\$X_5+X_1+X_2$}] at (0,{-\D}) {};
\end{scope}

\end{tikzpicture}
}%

\caption{$G_\text{sub}$ where there is one or two edges. The first two graphs are impossible for $\mais(G)=2$. For G0.2b, the length-2 index code shown here is also an index code for any $G$ (with five vertices) containing this graph.}
\label{fig:cat-1a}
\end{figure}

If there is no edge or only one edge, we can always find an induced subgraph of four vertices with no edge.  It follows from  Lemma~\ref{lem:4-cycle} that $\mais(G)\neq 2$ (contradiction).
Figure~\ref{fig:cat-1a} shows an example where the graph G0.1 contains only one edge $1 - 2$, and the subgraph induced by $\{2,3,4,5\}$ cannot contain any edge. 

Here, we use the notation G$x.y$, where $x$ is the length of the shortest undirected cycle in $G$, and $y$ is the number of edges. 

\subsubsection{There are only two edges} The two edges in $G$ can either be connected (see G0.2a in Figure~\ref{fig:cat-1a}) or disconnected (see G0.2b). We need to consider only non-isomorphic graphs, as the labelling of indices are arbitrary. 

For G0.2a, the subgraph induced by vertices $\{1,3,4,5\}$ contains no edge. By Lemma~\ref{lem:4-cycle}, this cannot happen. 

For G0.2b, since there is no edge in $\{1,4,5\}$, there must be a length-3 cycle. Without loss of generality (due to symmetry), let the cycle by $1 \rightarrow 4 \rightarrow 5 \rightarrow 1$. This necessitates the cycle in $\{1,3,5\}$ to be $1 \rightarrow 3 \rightarrow 5 \rightarrow 1$. The cycles in $\{2,3,5\}$ and $\{2,4,5\}$ must also take the forms shown in the figure.

Note that G0.2b is an interlinked cycle with \textit{inner vertices} $\{1,2,3,4\}$. The interlinked-cycle cover gives an index code of length 2: $[ (X_1+X_2+X_3+X_4)\,\, (X_5+X_1+X_2)]$ (see Definitions~\ref{definition:interlinked-cycles} and \ref{definition:ic-cover}). Here, it is understood that the addition is performed over the ring $\mathcal{X}$. 

So, any $G$ with 5 vertices, no undirected cycle, $\mais(G)=2$, and only two edges must contain an arc-deleted subgraph isomorphic to G0.2b. By Lemma~\ref{lem:remove-add-arcs}, $r_{m^t}(G) \leq r_{m^t}(\text{G0.2b}) \leq 2$. Since $2 = \mais(G) \leq r_{m^t}(G)$, we have $r_{m^t}(G)=2$.

\subsubsection{There are only three edges} 
\begin{figure}[t]
\centering
\resizebox{80mm}{!}{%
\begin{tikzpicture}
\linespread{1}
\def\D{7ex}
\graph [clockwise, n=5] {
{[path, edges=ultra thick]1, 2},
{[path, edges=ultra thick]1, 3},
{[path, edges=ultra thick]1, 4},
5
};
\node[label={[align=center]below:G0.3a\\impossible}] at (0,{-\D}) {};

\begin{scope}[xshift=40mm]
\graph [clockwise, n=5] {
{[path, edges=ultra thick]1, 2, 3, 4},
{[path, ->, edges={>=latex}] 5 -> 1 -> 3 -> 5}, 
{[path, ->, edges={>=latex}] 4 -> 5 -> 2 -> 4}, 
{[path, ->, edges={>=latex}]1-> 4}
};
\node[label={[align=center]below:G0.3b\\contains G0.2b}] at (0,{-\D}) {};
\end{scope}

\begin{scope}[xshift=80mm]
\graph [clockwise, n=5] {
{[path, edges={ultra thick}] 1, 2,3},
{[path, ->, edges={>=latex}] 3, 4},
{[path, edges={ultra thick}] 4, 5},
{[path, ->, edges={>=latex}] 5 -> 1 -> 3 -> 5}, 
{[path, ->, edges={>=latex}] 3 -> 4 -> 1}
};
\node[label={[align=center]below:G0.3c\\$X_1+X_2+X_3$\\$(X_4+X_5)+x_1$}] at (0,{-\D}) {};
\end{scope}

\end{tikzpicture}
}%

\caption{$G_\text{sub}$ where there are three edges and no undirected cycle. The first graph is impossible for $\mais(G)=2$, and there exists two-bit linear codes for the second and the third graphs.}
\label{fig:cat-1b}
\end{figure}
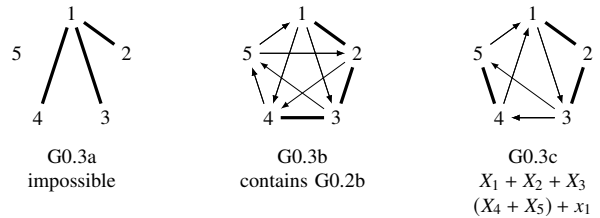

Without any undirected cycle, three edges can form only three non-isomorphic configurations as depicted in Figure~\ref{fig:cat-1b}.

If the three edges form a star, we have G0.3a. By Lemma~\ref{lem:4-cycle}, it is impossible as the subgraph induced by $\{2,3,4,5\}$ has no edge.

If the three edges form a path, we have G0.3b. The vertex set $\{1,4,5\}$ must contain a cycle. Without loss of generality (due to symmetry), let it be $1 \rightarrow 4 \rightarrow 5 \rightarrow 1$. The rest of the cycles for subgraphs with three vertices are then fixed. Since G0.3b contains G0.2b as an arc-deleted subgraph, invoking Lemma~\ref{lem:remove-add-arcs}, $r_{m^t}(\text{G0.3b}) \leq r_{m^t}(\text{G0.2b}) \leq 2$, and the length-2 index code for G0.2b also an index code for G0.3b.

If one of the three edges is disjoint from the other two, we have G0.3c. By symmetry and adding arcs to form cycles in $\{1,3,5\}$ and $\{1,3,4\}$, we have the configuration in the figure.  
G0.3c is an interlinked cycle with inner vertices $\{1,2,3\}$ and a super-vertex set $\{4,5\}$ (see Definition~\ref{def:super-vertex}). For an interlinked cycle of this type, the interlinked-cycle cover gives a index code $[(X_1 + X_2 + X_3)\,\, ((X_4+X_5) + X_1)]$ of length 2.

\subsubsection{There are only four edges} 
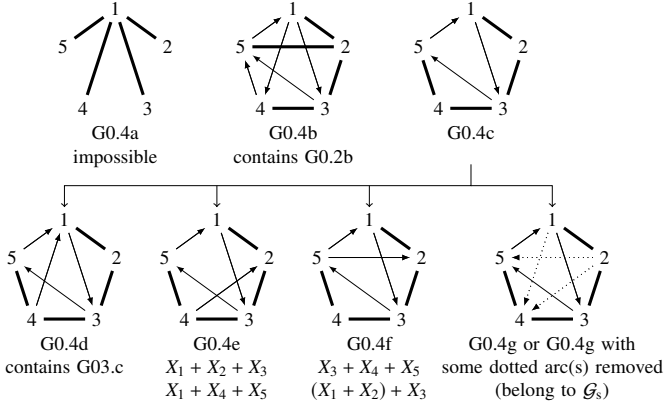
\begin{figure}[t]
\centering
\resizebox{90mm}{!}{%
  \begin{tikzpicture}
\linespread{1}
\def\D{7ex}
\def\X{\D*30/12}
\def\Y{\D*41.5/12}

\begin{scope}[xshift={\X/3}]
\graph [clockwise, n=5] {
{[path, edges=ultra thick]1, 2},
{[path, edges=ultra thick]1, 3},
{[path, edges=ultra thick]1, 4},
{[path, edges=ultra thick]1, 5}
};
\node[label={[align=center]below:G0.4a\\impossible}] at (0,{-0.8*\D}) {};
\end{scope}

\begin{scope}[xshift={1.5*\X}]
\graph [clockwise, n=5] {
{[path, edges=ultra thick]1, 2, 3, 4},
{[path, ->, edges={>=latex}] 5 -> 1 -> 3 -> 5}, 
{[path, edges=ultra thick] 5, 2},
{[path, ->, edges={>=latex}] 1 -> 4 -> 5}
};
\node[label={[align=center]below:G0.4b\\contains G0.2b}] at (0,{-0.8*\D}) {};
\end{scope}

\begin{scope}[xshift={8*\X/3}]
\graph [clockwise, n=5] {
{[path, edges={ultra thick}] 1, 2, 3, 4, 5},
{[path, ->, edges={>=latex}] 5 -> 1 -> 3 -> 5}, 
};
\node[label={[align=center]below:G0.4c}] at (0,{-0.8*\D}) {};
\end{scope}

\coordinate (a) at ({8*\X/3},{-1.65*\D});
\coordinate (b) at ({8*\X/3},{-2*\D});
\coordinate (c) at (0,{-2.35*\D});
\coordinate (d) at (\X,{-2.35*\D});
\coordinate (e) at ({2*\X},{-2.35*\D});
\coordinate (f) at ({3.2*\X},{-2.35*\D});

\draw[->] (a) -- (b) -| (c);
\draw[->] (b) -| (f);
\draw[<-] (d) -- (d.north |- b.west);
\draw[<-] (e) -- (e.north |- b.west);

\begin{scope}[xshift=0mm,yshift=-\Y]
\graph [clockwise, n=5] {
{[path, edges={ultra thick}] 1, 2, 3, 4, 5},
{[path, ->, edges={>=latex}] 5 -> 1 -> 3 -> 5}, 
{[path, ->, edges={>=latex}] 4 -> 1}
};
\node[label={[align=center]below:G0.4d\\contains G03.c}] at (0,{-0.8*\D}) {};
\end{scope}

\begin{scope}[xshift=\X,yshift=-\Y]
\graph [clockwise, n=5] {
{[path, edges={ultra thick}] 1, 2, 3, 4, 5},
{[path, ->, edges={>=latex}] 5 -> 1 -> 3 -> 5}, 
{[path, ->, edges={>=latex}] 4 -> 2}
};
\node[label={[align=center]below:G0.4e\\$X_1+X_2+X_3$\\$X_1+X_4+X_5$}] at (0,{-0.8*\D}) {};
\end{scope}

\begin{scope}[xshift={2*\X},yshift=-\Y]
\graph [clockwise, n=5] {
{[path, edges={ultra thick}] 1, 2, 3, 4, 5},
{[path, ->, edges={>=latex}] 5 -> 1 -> 3 -> 5}, 
{[path, ->, edges={>=latex}] 5 -> 2}
};
\node[label={[align=center]below:G0.4f\\$X_3+X_4+X_5$\\$(X_1+X_2)+X_3$}] at (0,{-0.8*\D}) {};
\end{scope}

\begin{scope}[xshift={3.2*\X},yshift=-\Y]
\graph [clockwise, n=5] {
{[path, edges={ultra thick}] 1, 2, 3, 4, 5},
{[path, ->, edges={>=latex}] 5 -> 1 -> 3 -> 5}, 
{[path, ->, edges={dotted,>=latex}] 1 -> 4},
{[path, ->, edges={dotted,>=latex}] 2 -> 4},
{[path, ->, edges={dotted,>=latex}] 2 -> 5}
};
\node[label={[align=center]below:G0.4g or G0.4g with\\ some dotted arc(s) removed\\(belong to $\mathcal{G}_\text{s}$)}] at (0,{-0.8*\D}) {};
\end{scope}

\end{tikzpicture}
}%
\caption{$G_\text{sub}$ where there are four edges and no undirected cycle.}
\label{fig:cat-1c}
\end{figure}

Without any undirected cycle, four edges can form only three non-isomorphic configurations G0.4a, G0.4b, or G0.4c in Figure~\ref{fig:cat-1c}.

\begin{figure}[t]
\centering
\resizebox{90mm}{!}{%
\begin{tikzpicture}
\linespread{1}
\def\D{7ex}
\def\X{\D*30/12}
\def\Y{\D*41.5/12}

\begin{scope}[xshift=0mm,yshift=0]
\graph [clockwise, n=5] {
{[clique, edges={ultra thick}] 1, 2, 3},
{[path, ->, edges={>=latex}] 4, 5, 1-> 4}, 
{[path, ->, edges={>=latex}] 5, 2, 4},
{[path, ->, edges={>=latex}] 5, 3, 4}
};
\node[label={[align=center]below:G3.3\\$X_4+X_4$\\$(X_1+X_2+X_3)+X_4$}] at (0,{-\D}) {};
\end{scope}

\begin{scope}[xshift=\X,yshift=0]
\graph [clockwise, n=5] {
{[clique, edges={ultra thick}] 1, 2, 3},
{[clique, edges={ultra thick, dashed}] 4, 5}
};
\node[label={[align=center]below:G3.4a\\$X_1+X_2+X_3$\\$X_4+X_5$}] at (0,{-\D}) {};
\end{scope}

\begin{scope}[xshift={2*\X},yshift=0]
\graph [clockwise, n=5] {
{[clique, edges={ultra thick}] 1, 2, 3},
{[path, ->, edges={>=latex}] 4, 5, 2-> 4}, 
{[path, ->, edges={>=latex}] 5, 3, 4},
{[clique, edges={ultra thick, dashed}] 1, 4}
};
\node[label={[align=center]below:G3.4b\\$X_1+X_2+X_3+X_4$\\$X_5+X_2+X_3$}] at (0,{-\D}) {};
\end{scope}

\begin{scope}[xshift={3.1*\X},yshift=0]
\graph [clockwise, n=5] {
{[clique, edges={ultra thick}] 1, 2, 3},
{[path, ->, edges={>=latex}] 4, 2, 5}, 
{[path, ->, edges={>=latex}] 5, 4, 3 -> 5},
{[clique, edges={ultra thick, dashed}] 1, 4}
};
\node[label={[align=center]below:G3.4c\\$X_1+X_2+X_3+X_4$\\$X_5+X_4$}] at (0,{-\D}) {};
\end{scope}


\begin{scope}[xshift=0mm,yshift=-\Y]
\graph [clockwise, n=5] {
{[clique, edges={ultra thick}] 1, 2, 3},
{[clique, edges={ultra thick, dashed}] 4, 3},
{[clique, edges={ultra thick, dashed}] 5, 1},
{[path, ->, edges={>=latex}] 4, 5, 2-> 4}, 
};
\node[label={[align=center]below:G3.5a\\contains G0.4e}] at (0,{-\D}) {};
\end{scope}

\begin{scope}[xshift=\X,yshift=-\Y]
\graph [clockwise, n=5] {
{[clique, edges={ultra thick}] 1, 2, 3},
{[clique, edges={ultra thick, dashed}] 4, 1},
{[clique, edges={ultra thick, dashed}] 5, 1},
{[path, ->, edges={>=latex}] 4, 5, 2-> 4}, 
{[path, ->, edges={>=latex}] 5, 3, 4} 
};
\node[label={[align=center]below:G3.5b\\contains G3.4b}] at (0,{-\D}) {};
\end{scope}

\begin{scope}[xshift={2*\X},yshift=-\Y]
\graph [clockwise, n=5] {
{[clique, edges={ultra thick}] 1, 2, 3},
{[path, ->, edges={>=latex}] 4, 5, 3-> 4}, 
{[clique, edges={ultra thick, dashed}] 5, 1},
{[clique, edges={ultra thick, dashed}] 5, 2}
};
\node[label={[align=center]below:G3.5c\\$X_1+X_2+X_3+X_5$\\$X_4+X_5$}] at (0,{-\D}) {};
\end{scope}


\begin{scope}[xshift=0mm,yshift={-2*\Y}]
\graph [clockwise, n=5] {
{[clique, edges={ultra thick}] 1, 2, 3},
{[clique, edges={ultra thick, dashed}] 4, 1},
{[clique, edges={ultra thick, dashed}] 4, 2},
{[clique, edges={ultra thick, dashed}] 4, 3},
5
};
\node[label={[align=center]below:G3.6a\\$X_1+X_2+X_3+X_4$\\$X_5$}] at (0,{-\D}) {};
\end{scope}

\begin{scope}[xshift=\X,yshift={-2*\Y}]
\graph [clockwise, n=5] {
{[clique, edges={ultra thick}] 1, 2, 3},
{[clique, edges={ultra thick, dashed}] 4, 2},
{[clique, edges={ultra thick, dashed}] 4, 3},
{[clique, edges={ultra thick, dashed}] 5, 1}
};
\node[label={[align=center]below:G3.6b\\contains G3.4a}] at (0,{-\D}) {};
\end{scope}

\begin{scope}[xshift={2*\X},yshift={-2*\Y}]
\graph [clockwise, n=5] {
{[clique, edges={ultra thick}] 1, 2, 3},
{[clique, edges={ultra thick, dashed}] 4, 2},
{[clique, edges={ultra thick, dashed}] 4, 3},
{[clique, edges={ultra thick, dashed}] 5, 2},
{[path, ->, edges={>=latex}] 5 -> 1 -> 4 -> 5} 
};
\node[label={[align=center]below:G3.6c\\contains G3.4c}] at (0,{-\D}) {};
\end{scope}

\begin{scope}[xshift={3*\X},yshift={-2*\Y}]
\graph [clockwise, n=5] {
{[clique, edges={ultra thick}] 1, 2, 3},
{[clique, edges={ultra thick, dashed}] 4, 2},
{[clique, edges={ultra thick, dashed}] 4, 3},
{[clique, edges={ultra thick, dashed}] 5, 2},
{[path, ->, edges={>=latex}] 5 -> 4 -> 1 -> 5} 
};
\node[label={[align=center]below:G3.6d\\contains G3.4c}] at (0,{-\D}) {};
\end{scope}

\end{tikzpicture}
}%
\caption{$G_\text{sub}$ where there is a length-3 undirected cycle (among vertices 1, 2, and 3; marked with solid lines). Additional edges are marked with dashed lines. Arcs are then added so that every three vertices must contain at least one cycle.}
\label{fig:cat-2}
\end{figure}
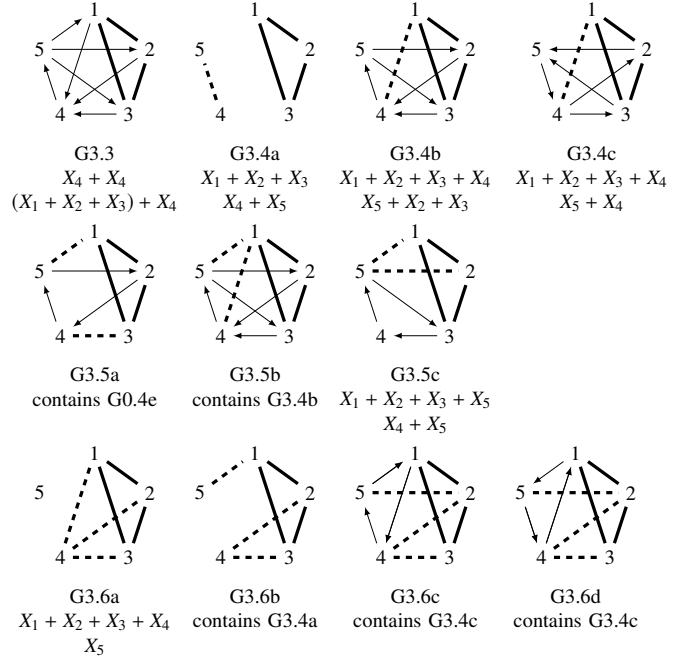

If the four edges form a star, i.e., G0.4a, it is an impossible subgraph as $\{2,3,4,5\}$ does not contain any edge.

For configuration G0.4b, the vertex set $\{1,4,5\}$ must contain a length-3 cycle. Without loss of generality (due to symmetric), let an arc in the cycle be $5 \rightarrow 1$, and so the cycle is $1 \rightarrow 4 \rightarrow 5 \rightarrow 1$. With this, the cycles for $\{1,3,5\}$ is also fixed. We see that this graph contains G0.2b as an arc-deleted subgraph, and hence
$r_{m^t}(\text{G0.4b}) \leq r_{m^t}(\text{G0.2b}) \leq 2$, and the length-2 linear code for G0.2b is also an index code for G0.4b.

If the four edges form a path, 
 we need to further categorise all $G$ that contain G0.4c. Since the positions of edges in G0.4c are fixed, and we can only add arcs. The only positions to add arcs are within the pairs $\{(1,4), (2,4), (2,5)\}$, and we can only add at most one arc in each pair (adding arcs in both directions forms an edge). So, any $G$ in this category must satisfy either of the following:
\begin{itemize}
\item If there is an additional arc within any pair in $\{(1,4), (2,4), (2,5)\}$ from a larger index to a smaller index, i.e., $4 \rightarrow 1$, $4 \rightarrow 2$, or $5 \rightarrow 2$, we get a graph that contains  G0.4d, G0.4e, or G0.4f, respectively, as an arc-deleted subgraph. Note that a graph can also simultaneously contain more than one of these graphs as subgraphs. Note that
\begin{itemize}
\item G0.4d contains G0.3c as a subgraph;
\item G0.4e has a length-2 index code $[(X_1+X_2+X_3)\,\,(X_1+X_4+X_5)]$ ;
\item G0.4f contains an interlinked cycle with inner vertices $\{3,4,5\}$ and a super-vertex set $\{1,2\}$.
\end{itemize}
\item Otherwise, we must get G0.4g or G0.4g with some of the dotted arcs removed. These graphs belong to $\mathcal{G}_\text{s}$, and  we will deal it Theorem~\ref{thm:5-others}.
\end{itemize}

A length-2 linear code exists for each of G0.4d, -e, or -f. 
\subsubsection{There are five of more edges} This configuration is impossible as it is known to contain an undirected cycle. 

So, we have shown that for any $G \notin \mathcal{G}_\text{s}$ such that $|V(G)|=5$, $\mais(G)=2$, and $G$ contains no undirected cycle, then it must contain either G0.2b, G0.3c, G0.4e, or G0.4f as an arc-deleted subgraph.
For any case, $r(G) = r_{m^t}(G) = 2$ for any $m$ and $t$.

\subsection{Category 2: An undirected cycle of length~3}

Without loss of generality, let the undirected cycle be $1 - 2 - 3 - 1$ (depicted as solid lines in Figure~\ref{fig:cat-2}). First, if there is an additional edge $4 - 5$ (denoted by G3.4a in Figure~\ref{fig:cat-2}), there exists a length-2 index code using the clique cover $[(X_1+X_2)\,\,(X_3+X_4+X_5)]$. 

Otherwise (i.e., no edge between 4 and 5), any additional edge (in addition to $1 - 2 - 3 - 1$) must be between  $\{1,2,3\}$ and $\{4,5\}$. For this, we have the following categories, grouped by the number of additional edge (dashed lines in Figure~\ref{fig:cat-2}):

\subsubsection{No edge between the groups $\{1,2,3\}$ and $\{4,5\}$} The only non-isomorphic graph where every three vertices contain a cycle is depicted in G3.3. This is an interlinked cycle with inner vertices $\{4,5\}$ and a super-vertex set $\{1,2,3\}$. An index code for this graph is $[(X_4+X_5)\,\,((X_1+X_2+X_3)+X_4)]$.

\subsubsection{One edge between the groups} Without loss of generality, let the additional edge be $1 - 4$. Two non-isomorphic graphs with different arc positions are possible: G3.4b and G3.4c. They are interlinked cycles with inner vertices $\{1,2,3,4\}$.

\subsubsection{Two edges between the groups} If the two edges connect four different vertices, we have G3.5a. If the two edges connect between the same vertex in $\{1,2,3\}$ to two different vertices in $\{4,5\}$, we have G3.5b. Otherwise, if the two edges connect between different vertices in $\{1,2,3\}$ to the same vertex in $\{4,5\}$, we have G3.5c, which is an interlinked cycle with inner vertices $\{1,2,3,5\}$.

\subsubsection{Three edges between the groups} The three edges can be placed in three non-isomorphic positions: (i) Between three vertices in $\{1,2,3\}$ and one vertex in $\{4,5\}$, we have G3.6a; (ii) Between three vertices in $\{1,2,3\}$ and two vertices in $\{4,5\}$, we have G3.6b; (iii) Between two vertices in $\{1,2,3\}$ and two vertices in $\{4,5\}$, we have G3.6c and G3.6d. For G3.6a, the clique cover gives a linear index code $[(X_1+X_2+X_3+X_4)\,\,X_5]$.

\subsubsection{Four or more edges between the groups} We can show that the graph will always contain G3.4a with vertex relabelling.

So, we have shown that for any $G \notin \mathcal{G}_\text{s}$ such that $|V(G)|=5$, $\mais(G)=2$, and $G$ contains an undirected cycle of length 3, then there exists a linear index code of length 2, which can be constructed using the interlinked-cycle cover (which includes the clique cover as a special case). 

\subsection{Category 3: An undirected cycle of length~4 and no undirected cycle of length~3}

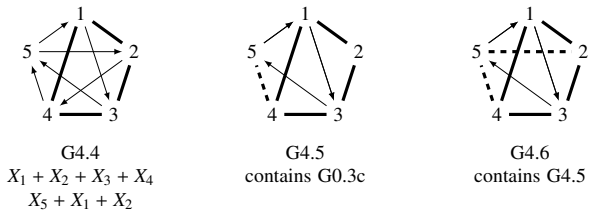
\begin{figure}[t]
\centering
\resizebox{80mm}{!}{%
\begin{tikzpicture}
\linespread{1}
\def\D{7ex}
\graph [clockwise, n=5] {
{[path, edges=ultra thick] 1, 2, 3, 4 -- 1},
{[path, ->, edges={>=latex}] 4, 5, 1, 3},
{[path, ->, edges={>=latex}] 3, 5, 2, 4} 
};
\node[label={[align=center]below:G4.4\\$X_1+X_2+X_3+X_4$\\$X_5+X_1+X_2$}] at (0,{-\D}) {};

\begin{scope}[xshift=40mm]
\graph [clockwise, n=5] {
{[path, edges=ultra thick] 1, 2, 3, 4 -- 1},
{[clique, edges={ultra thick, dashed}] 4, 5},
{[path, ->, edges={>=latex}] 5 -> 1 -> 3 -> 5} 
};
\node[label={[align=center]below:G4.5\\contains G0.3c}] at (0,{-\D}) {};
\end{scope}

\begin{scope}[xshift=80mm]
\graph [clockwise, n=5] {
{[path, edges=ultra thick] 1, 2, 3, 4 -- 1},
{[path, edges={ultra thick, dashed}] 4, 5, 2},
{[path, ->, edges={>=latex}] 5 -> 1 -> 3 -> 5}
};
\node[label={[align=center]below:G4.6\\contains G4.5}] at (0,{-\D}) {};
\end{scope}

\end{tikzpicture}
}%
\caption{$G_\text{sub}$ where there is a length-4 undirected cycle (marked with thick solid lines) and no length-3 undirected cycle. Additional edges are marked with dashed lines. Arcs are then added so that every three vertices must contain at least one cycle. There are only three non-isomorphic graphs.}
\label{fig:cat-3}
\end{figure}

Next, we consider the category where there is an undirected cycle of length~4; without loss of generality, let the cycle be $1 - 2 - 3 - 4 - 1$. We find graphs when there is (i) no additional edge, (ii) one additional edge, or (iii) two additional edges. Note that there cannot be three additional edge, as it will create a length-3 undirected cycle. For each graph here, there exists a length-2 linear index code, as shown in Figure~\ref{fig:cat-3}. Note that G4.4 is an interlinked cycle with inner vertices $\{1,2,3,4\}$.

\subsection{Category 4: An undirected cycle of length~5 and no undirected cycle of length~3 or 4}

Without loss of generality, let the undirected cycle be $1 - 3 - 5 - 2 - 4 - 1$. With this, there cannot be any additional edge; otherwise, we get a length-3 or -4 cycle. Also, any additional arc must be between adjacent vertices on the ``circumference'', i.e., within any pair in $\{(1,2), (2,3), (3,4), (4,5), (5,1)\}$.

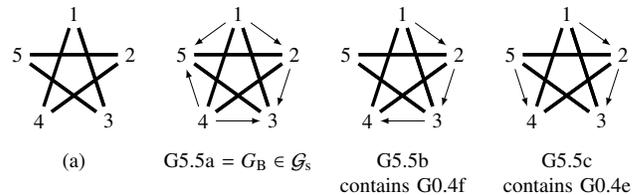
\begin{figure}[t]
\centering

\resizebox{85mm}{!}{%
\begin{tikzpicture}
\linespread{1}
\def\D{7ex}
\def\X{\D*30/12}
\def\Y{\D*41.5/12}

\begin{scope}[xshift=0mm,yshift=0]
\graph [clockwise, n=5] {
1,2,3,4,5,
{[path, edges={ultra thick}] 1 -- 3 -- 5 -- 2 -- 4 -- 1}
};
\node[label={[align=center]below:(a)}] at (0,{-\D}) {};
\end{scope}

\begin{scope}[xshift=\X,yshift=0]
\graph [clockwise, n=5] {
1,2,3,4,5,
{[path, edges={ultra thick}] 1 -- 3 -- 5 -- 2 -- 4 -- 1},
{[edges={>=latex}]1 -> 2 -> 3 <- 4 -> 5 <- 1 },
};
\node[label={[align=center]below:G5.5a $=G_\text{B} \in \mathcal{G}_\text{s}$}] at (0,{-\D}) {};
\end{scope}

\begin{scope}[xshift={2*\X},yshift=0]
\graph [clockwise, n=5] {
1,2,3,4,5,
{[path, edges={ultra thick}] 1 -- 3 -- 5 -- 2 -- 4 -- 1},
{[edges={>=latex}]1 -> 2 -> 3 -> 4},
};
\node[label={[align=center]below:G5.5b\\contains G0.4f}] at (0,{-\D}) {};
\end{scope}

\begin{scope}[xshift={3*\X},yshift=0]
\graph [clockwise, n=5] {
1,2,3,4,5,
{[path, edges={ultra thick}] 1 -- 3 -- 5 -- 2 -- 4 -- 1},
{[edges={>=latex}]1 -> 2 -> 3, 5 -> 4 },
};
\node[label={[align=center]below:G5.5c\\contains G0.4e}] at (0,{-\D}) {};
\end{scope}
\end{tikzpicture}
}%
\caption{$G_\text{sub}$ where there is a length-5 undirected cycle (marked with thick lines) and no length-3 or -4 undirected cycle.}
\label{fig:cat-4}
\end{figure}

If we add arcs in the way to obtain G5.5a, we get a graph in $\mathcal{G}_\text{s}$. We will deal with this in the next section.

We now show that for any graph in Category 4 that is not a subgraph of G5.5a, there exists a two-bit linear index code. First, if we add (i) zero, (ii) one, or (iii) two arcs to Figure~\ref{fig:cat-4}(a), we must get an isomorphic arc-deleted subgraph of G5.5a, and they are members in $\mathcal{G}_\text{s}$.

 If we add three arcs, the only graphs that are not isomorphic arc-deleted subgraphs of G5.5a are G5.5b and G5.5c. By relabelling the vertices, G5.5b contains G0.4f, and G5.5c contains G0.4e.

If we add four arcs to Figure~\ref{fig:cat-4}(a), they must form a string (i.e., a path where the direction of the arcs can be arbitrary) on the circumference (dashed lines on Figure~\ref{fig:cat-4}(a)). The only non-isomorphic combinations of length-4 strings along the circumference are: 
(i) $\rightarrow \rightarrow \rightarrow \rightarrow$, 
(ii) $\rightarrow \rightarrow \rightarrow \leftarrow$,
(iii) $\leftarrow \leftarrow \leftarrow \rightarrow$, 
(iv) $\rightarrow \rightarrow \leftarrow \leftarrow$, 
(v) $\leftarrow \leftarrow \rightarrow \rightarrow$,
(vi) $\rightarrow \rightarrow \leftarrow \rightarrow$,
(vii) $\leftarrow \leftarrow \rightarrow \leftarrow$, 
(viii) $\rightarrow \leftarrow \leftarrow \rightarrow$,
(ix) $\rightarrow \leftarrow \rightarrow \leftarrow$, and
(x) $\leftarrow \rightarrow \leftarrow \rightarrow$. 
Configurations (i)--(iii) each contain G5.5b, (iv)--(v) each contain G5.5c, (vi)--(x) each are subgraphs of G5.5a (i.e., members of $\mathcal{G}_\text{s}$).

Lastly, we add five arcs, i.e., one arc within any pair in $\{(1,2), (2,3), (3,4), (4,5), (5,1)\}$. We will now show that the graph must be G5.5a, G5.5b$^+$, or G5.5c$^+$.\footnote{Recall that $G^+$ contains $G$ as an arc-deleted subgraph.} We can easily show that there must be a two adjacent arc in the same direction. Without loss of generality, let them be $1 \rightarrow 2 \rightarrow 3$.
For arcs between $\{3,4\}$, $\{4,5\}$, and $\{1,5\}$, if any of them does not follow the direction as that in G5.5a, we have either G5.5b$^+$ or G5.5c$^+$.


We have shown that for any $G \notin \mathcal{G}_\text{s}$ such that $|V(G)|=5$, $\mais(G)=2$, and $G$ contains an undirected cycle of length~5, and no undirected cycle of length~3 or 4, then it must contain G5.5b or G5.5c as an arc-deleted subgraph. So, $r(G) = r_{m^t}(G) = 2$.

This completes the proof of Lemma~\ref{proposition:last-case}.  $\hfill \blacksquare$

\begin{remark}
For all graph $G \notin \mathcal{G}_\text{s}$ such that $|V(G)|=5$ and $\mais(G)=2$, except those that contain G0.4e, an optimal scalar linear index code of length~2 can be constructed using the interlinked-cycle cover.
\end{remark}

\section{Proof of Theorem~\ref{thm:5-others}} \label{theorem-proof}

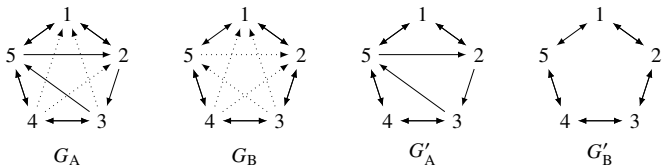
\begin{figure}[t]
\centering
\resizebox{90mm}{!}{%
\begin{tikzpicture}
\def\D{18mm}
\def\X{30mm}
\def\Y{41.5mm}

\begin{scope}[xshift=0mm,yshift=0]
\graph [clockwise, n=5] {
1,2,3,4,5,
{[path,<->, edges={>=latex}] 1  <-> 2},
{[path,<->, edges={>=latex}] 3  <-> 4  <-> 5  <-> 1},
{[edges={>=latex}] 2 -> 3, 3 -> 5, 5 -> 2 },
{[edges={dotted,>=latex}] 3 -> 1, 4 -> 1, 4 -> 2}
};
\node[label={[align=center]$G_\text{A}$}] at (0,{-\D}) {};
\end{scope}

\begin{scope}[xshift=\X,yshift=0]
\graph [clockwise, n=5] {
1,2,3,4,5,
{[path,<->, edges={>=latex}] 1  <-> 2  <-> 3  <-> 4  <-> 5  <-> 1},
{[edges={dotted,>=latex}] 5 -> 2, 4 -> 1, 4 -> 2, 3 -> 1, 3 -> 5 }
};
\node[label={[align=center]$G_\text{B}$}] at (0,{-\D}) {};
\end{scope}

\begin{scope}[xshift=2*\X,yshift=0]
\graph [clockwise, n=5] {
1,2,3,4,5,
{[path,<->, edges={>=latex}] 1  <-> 2},
{[path,<->, edges={>=latex}] 3  <-> 4  <-> 5  <-> 1},
{[edges={>=latex}] 2 -> 3,  3 -> 5,  5 -> 2 }
};
\node[label={[align=center]$G_\text{A}'$}] at (0,{-\D}) {};
\end{scope}

\begin{scope}[xshift=3*\X,yshift=0]
\graph [clockwise, n=5] {
1,2,3,4,5,
{[path,<->, edges={>=latex}] 1  <-> 2  <-> 3  <-> 4  <-> 5  <-> 1},
};
\node[label={[align=center]$G_\text{B}'$}] at (0,{-\D}) {};
\end{scope}

\end{tikzpicture}
}%
\caption{$G_\text{A}$ and $G_\text{B}$ with vertices labelled.}
\label{fig:graphs-submod}
\end{figure}

Refer to $G_\text{A}$ and $G_\text{B}$ in Figure~\ref{fig:special-group}.
Denote $G'_\text{A}$ and $G'_\text{B}$ as the subgraphs formed by removed all dotted arcs in $G_\text{A}$ and $G_\text{B}$ respectively. Blasiak et al.~\cite{blasiakkleinberglubertzky13} found $r(G)$ for all undirected cycles, which include the 5-cycle $G'_\text{B}$ as a special case. Here, we need to further find $r_{m^t}(G)$ for all $G \in \mathcal{G}_\text{s}$.

We first label the vertices of $G_\text{A}$, $G_\text{B}$, $G_\text{A}'$, and $G_\text{B}'$ as in Figure~\ref{fig:graphs-submod}.

\textit{(Achievability):} 
Let each message be a vector of length~2, which can be written as $X_i = (X_i^{(1)}, X_i^{(2)}) \in \mathcal{X}' \times \mathcal{X}'$, where $\mathcal{X}'= \{0,1,\dotsc,m^k-1\}$. Let $\mathcal{Y} = (\mathcal{X}')^5$, $\phi = [\phi_1 \dotsm \phi_5]$ such that $\phi_i: \mathcal{X}'^{2 \times 5} \mapsto \mathcal{X}'$ is vector linear over the ring $\mathcal{X}'$. The codelength here is 2.5.

For the graph $G'_\text{B}$, by time-sharing the cycle cover over the cycles $\{(i,i+1 \mod 5): i \in [5]\}$, we obtain the following index code: $\phi_1 = X_1^{(1)} + X_2^{(1)}$, $\phi_2 = X_2^{(2)} + X_3^{(2)}$, $\phi_3 = X_3^{(1)} + X_4^{(1)}$, $\phi_4 = X_4^{(2)} + X_5^{(2)}$, $\phi_5 = X_5^{(1)} + X_1^{(2)}$, where the addition is performed over modulo-$m^k$. 

For the graph $G'_\text{A}$, by time-sharing the index codes for the interlinked cycle $\{1,2,3,5\}$ (with inner vertices $\{1,2,3\}$) and cycles $\{1,2\}$, $\{3,4\}$, $\{4,5\}$, we obtain the following index code:
$\phi_1 = X_1^{(1)} + X_2^{(1)} + X_3^{(1)}$, $\phi_2 = X_5^{(1)} + X_1^{(1)}$, $\phi_3 = X_1^{(2)} + X_2^{(2)}$, $\phi_4 = X_3^{(2)} + X_4^{(1)}$, $\phi_5 = X_4^{(2)} + X_5^{(2)}$. 

Note that any $G \in \mathcal{G}_\text{s}$ must contain either $G_\text{A}'$ or $G_\text{B}'$ as an arc-deleted subgraph. Invoking Lemma~\ref{lem:remove-add-arcs}, we have that 
$r_{m^{2k}}(G) \leq 2.5$, and the upper bounds can be attained by vector linear codes over the ring $\mathcal{X}'$.

\textit{(Lower bound):} 
While, upper bounds found for $G'_\text{A}$ and $G'_\text{B}$ is applicable to all $G \in \mathcal{G}_\text{s}$, for lower bounds, we need to consider $G_\text{A}$ and $G_\text{B}$.

We will now use the following tools to find lower bounds for $G_\text{A}$ and $G_\text{B}$:
\begin{enumerate}
\item\label{item:a} \textit{Submodularity of entropy:} Entropy is a submodular function, i.e., for any sets of random variables $S$ and $T$, we must have that
\begin{equation}
H(S) + H(T) \geq H(S \cup T) + H(S \cap T). \label{eq:submod}
\end{equation}
\item\label{item:b} \textit{Decodability:} Given $G$. For any vertex~$i \in V(G)$ with out-neighbourhood $\on{G}{i}$, receiver~$i$ must be able to decode $X_i$ given the index code, denoted by $Y$, and all the messages it knows a priori, $\bm{X}_{\on{G}{i}}$, i.e.,
\begin{align}
&&H(X_i|Y,\bm{X}_{\on{G}{i}}) &= 0 & \\
 \Rightarrow &&H(\bm{X}_{\on{G}{i}},Y) &= H(\bm{X}_{\{i\} \cup \on{G}{i}},Y) & \label{eq:decod}
\end{align}
\end{enumerate}
Note that while the submodularity inequality~\eqref{eq:submod} is universal in the sense it does not depend on specific graphs, the decodability equality~\eqref{eq:decod} does depend on $G$.

We now derive submodularity and decodability conditions can be applied to both $G_\text{A}$ and $G_\text{B}$, which are based on those for undirected cycles~\cite{blasiakkleinberglubertzky13}. Let $q = \log_2|\mathcal{X}|$. 

\ifx\doublecolumn\undefined
\begin{align}
 H(Y) + 2q &= H(Y) + H(\bm{X}_{\{2,5\}}) \geq H(\bm{X}_{\{2,5\}},Y) \label{eq:submod-1}\\
 H(Y) + 2q &= H(Y) + H(\bm{X}_{\{1,3\}}) \geq H(\bm{X}_{\{1,3\}},Y) \label{eq:submod-2}\\ 
H(Y) + q &= H(Y) + H(X_4) \geq H(X_4,Y) \label{eq:submod-3}\\
H(\bm{X}_{\{1,2,5\}},Y) + H(\bm{X}_{\{1,2,3\}},Y)  &\geq H(\bm{X}_{\{1,2,3,5\}}, Y) + H(\bm{X}_{\{1,2\}},Y) \label{eq:submod-4} \\
& = H(\bm{X}_{\{1,2,3,4,5\}}, Y) + H(\bm{X}_{\{1,2\}},Y) \label{eq:decod-1}\\
H(\bm{X}_{\{1,2\}},Y) + H(X_4,Y) 
& \geq H(\bm{X}_{\{1,2,4\}}, Y) + H(Y) \label{eq:submod-5} \\
& = H(\bm{X}_{\{1,2,4,5\}}, Y) + H(Y) \label{eq:decod-2}\\
&  = H(\bm{X}_{\{1,2,3,4,5\}}, Y) + H(Y) \label{eq:decod-3}\\
H(\bm{X}_{\{2,5\}},Y) &= H(\bm{X}_{\{1,2,5\}},Y) \label{eq:decod-4}\\
H(\bm{X}_{\{1,3\}},Y) &= H(\bm{X}_{\{1,2,3\}},Y) \label{eq:decod-5}
\end{align}
\else
\begin{align}
& H(Y) + 2q = H(Y) + H(\bm{X}_{\{2,5\}}) \geq H(\bm{X}_{\{2,5\}},Y) \label{eq:submod-1}\\
& H(Y) + 2q = H(Y) + H(\bm{X}_{\{1,3\}}) \geq H(\bm{X}_{\{1,3\}},Y) \label{eq:submod-2}\\ 
& H(Y) + q = H(Y) + H(X_4) \geq H(X_4,Y) \label{eq:submod-3}\\
& H(\bm{X}_{\{1,2,5\}},Y) + H(\bm{X}_{\{1,2,3\}},Y) \notag \\
& \quad \geq H(\bm{X}_{\{1,2,3,5\}}, Y) + H(\bm{X}_{\{1,2\}},Y) \label{eq:submod-4} \\
& \quad = H(\bm{X}_{\{1,2,3,4,5\}}, Y) + H(\bm{X}_{\{1,2\}},Y) \label{eq:decod-1}\\
& H(\bm{X}_{\{1,2\}},Y) + H(X_4,Y) \notag \\
& \quad  \geq H(\bm{X}_{\{1,2,4\}}, Y) + H(Y) \label{eq:submod-5} \\
& \quad = H(\bm{X}_{\{1,2,4,5\}}, Y) + H(Y) \label{eq:decod-2}\\
& \quad = H(\bm{X}_{\{1,2,3,4,5\}}, Y) + H(Y) \label{eq:decod-3}\\
& H(\bm{X}_{\{2,5\}},Y) = H(\bm{X}_{\{1,2,5\}},Y) \label{eq:decod-4}\\
& H(\bm{X}_{\{1,3\}},Y) = H(\bm{X}_{\{1,2,3\}},Y) \label{eq:decod-5}
\end{align}
\fi

Here, inequalities \eqref{eq:submod-1}--\eqref{eq:submod-4}, \eqref{eq:submod-5} follow from submodularity of entropy, and equalities \eqref{eq:decod-1}, \eqref{eq:decod-2}--\eqref{eq:decod-5} from decodability.

Now,
\ifx\doublecolumn\undefined
\begin{subequations}
\begin{align}
10q &= 2 H(\bm{X}_{[5]}) = 2 H(\bm{X}_{[5]},Y) \notag \\
& \leq
(  H(\bm{X}_{\{1,2,5\}},Y) + H(\bm{X}_{\{1,2,3\}},Y) - H(\bm{X}_{\{1,2\}},Y) ) + ( H(\bm{X}_{\{1,2\}},Y) + H(X_4,Y) - H(Y) ) \label{eq:vector-upper-1}\\
& = H(\bm{X}_{\{2,5\}},Y) + H(\bm{X}_{\{1,3\}},Y) + H(X_4,Y) - H(Y)   \label{eq:vector-upper-2} \\
& \leq H(Y) + 2q + H(Y) + 2q + H(Y) + q + - H(Y), \label{eq:vector-upper-3}
\end{align}
\end{subequations}
\else
\begin{subequations}
\begin{align}
& 10q = 2 H(\bm{X}_{[5]}) = 2 H(\bm{X}_{[5]},Y) \notag \\
& \leq
(  H(\bm{X}_{\{1,2,5\}},Y) + H(\bm{X}_{\{1,2,3\}},Y) - H(\bm{X}_{\{1,2\}},Y) ) \notag \\
& \quad + ( H(\bm{X}_{\{1,2\}},Y) + H(X_4,Y) - H(Y) ) \label{eq:vector-upper-1}\\
& = H(\bm{X}_{\{2,5\}},Y) + H(\bm{X}_{\{1,3\}},Y) + H(X_4,Y) - H(Y)   \label{eq:vector-upper-2} \\
& \leq H(Y) + 2q + H(Y) + 2q + H(Y) + q + - H(Y), \label{eq:vector-upper-3}
\end{align}
\end{subequations}
\fi
where $[5] \triangleq \{1,2,\dotsc,5\}$, \eqref{eq:vector-upper-1} follows from \eqref{eq:decod-1}--\eqref{eq:decod-3};
\eqref{eq:vector-upper-2} follows from \eqref{eq:decod-4}--\eqref{eq:decod-5}; 
\eqref{eq:vector-upper-3} follows from \eqref{eq:submod-1}--\eqref{eq:submod-3}. This gives
\begin{equation}
p \log_2 |\mathcal{Y}| \geq H(Y) \geq 2.5q = 2.5 \log_2 |\mathcal{X}|,
\end{equation}
for any index code $C$, and hence $r_{m^{2k}}(G_\text{A}) \geq 2.5$, and  $r_{m^{2k}}(G_\text{B}) \geq 2.5$.

Note that any $G \in \mathcal{G}_\text{s}$ must be either $G_\text{A}$ or $G_\text{B}$, or an arc-deleted subgraph of either of them. Invoking Lemma~\ref{lem:remove-add-arcs}, we have that 
$r_{m^{2k}}(G) \geq 2.5$.

Combining this lower bound and the achievability results, we have $r_{m^{2k}}(G) = 2.5$ for all $m$ and $k$, and hence $r(G) = 2.5$.

Now, note that $\mais(G) \leq \mais(G^-)$, as removing arcs can only reduce the number of cycles. We can manually verify that $\mais(G_\text{A}) = \mais(G_\text{B}) = \mais(G_\text{A}') = \mais(G_\text{B}') = 2$. Since any $G \in \mathcal{G}_\text{s}$ must satisfy (i) $G = (G_\text{A})^-$ or $G=(G_\text{B})^-$, and (ii) $G= (G_\text{A}')^+$ or $G=(G_\text{B}')^+$, we have
$\mais(G) =2$.

This completes the proof of Theorem~\ref{thm:5-others}. $\hfill \blacksquare$

\end{document}